\documentclass[twoside]{JHEP3} 
\usepackage{booktabs}

\usepackage{epsfig,multicol,multirow,subfigure,rotating}
\usepackage{graphicx}
\usepackage[latin1]{inputenc}
\usepackage{amssymb}
\usepackage{amsmath}
\usepackage{epsfig}
\newcommand\fverb{\setbox\pippobox=\hbox\bgroup\verb}
\newcommand\fverbdo{\egroup\medskip\noindent%
                      \fbox{\unhbox\pippobox}\ }

\newcommand\fverbit{\egroup\item[\fbox{\unhbox\pippobox}]}
\newbox\pippobox
\newcommand{\ie}{{\it i.e.}}

\newcommand{\be}{\begin{equation}}
\newcommand{\ee}{\end{equation}}
\newcommand{\ba}{\begin{eqnarray}}
\newcommand{\ea}{\end{eqnarray}}
\newcommand{\bt}{\begin{tabular}}
\newcommand{\et}{\end{tabular}}
\newcommand{\bfig}{\begin{figure}}
\newcommand{\efig}{\end{figure}}

\newcommand\sss{\scriptscriptstyle\rm}

\newcommand\muR{\mu_{\sss R}}
\newcommand\muF{\mu_{\sss F}}

\title{NLO predictions for t-channel production of single top and fourth generation 
quarks at hadron colliders}

\author{John M. Campbell\\
Department of Physics and Astronomy, University of Glasgow \\
Glasgow G12 8QQ, United Kingdom\\
E-mail: \email{j.campbell@physics.gla.ac.uk}
}

\author{Rikkert Frederix\footnote{On leave of absence from CP3, Universit\'{e} catholique de Louvain.} \\
PH Department, Theory group, CERN \\
1211-CH Geneva, Switzerland\\
E-mail: \email{rikkert.frederix@cern.ch}
}

\author{Fabio Maltoni \\
Centre for Particle Physics and Phenomenology (CP3) \\
Universit\'{e} catholique de Louvain\\
Chemin du Cyclotron 2, B-1348 Louvain-la-Neuve, Belgium\\
E-mail: \email{fabio.maltoni@uclouvain.be}
}

\author{Francesco Tramontano\\
Universit\`a di Napoli Federico II\\
Dipartimento di Scienze Fisiche, and INFN, Sezione di Napoli\\
I-80126 Napoli, Italy\\
E-mail: \email{francesco.tramontano@na.infn.it}
}

\preprint{CERN-PH-TH/2009-129\\CP3-09-31\\DSF-11/09}            

\abstract{We present updated NLO predictions for the electroweak
  $t$-channel production of heavy quarks at the Tevatron and at the
  LHC. We consider production of single top and fourth generation
  $t^\prime$ starting from both $2 \to 2$ and $2 \to 3$ Born
  processes. Predictions for $tb'$ and $t^\prime b'$ cross sections at
  NLO are also given for the first time. A thorough study of the
  theoretical uncertainties coming from parton distribution functions,
  renormalisation and factorisation scale dependence and heavy quark
  masses is performed.}


\begin{document}

\section{Introduction}
\label{sec:intro}

The recent observation of electroweak production of single top at the
Tevatron~\cite{Aaltonen:2009jj,Abazov:2009ii} has opened a new
exciting chapter in top physics. Many other properties of the heaviest
of the known quarks are now accessible and will be studied over the
coming years at both the Tevatron and the LHC.  For instance, single
top production has already provided the first direct measurement of
$V_{tb}$, \ie~without appealing to unitarity of the CKM
matrix.\footnote{$V_{tb}\gg V_{ts},V_{td}$, however, has been assumed
  in the current analyses~\cite{Aaltonen:2009jj,Abazov:2009ii}.  Such
  a constraint could also be lifted with further integrated
  luminosity~\cite{Alwall:2006bx}.} In fact a significant motivation to
accurately measure cross sections and distributions for this final
state is given by the search for new physics
effects~\cite{Tait:2000sh}.  Flavor changing neutral currents, the
existence of $W^{\prime\pm}$, $H^\pm$ or of more general four-fermion
interactions could be detected as anomalous production of single top.
In addition, production of a heavier partner of the top or bottom
quarks -- associated with a fourth generation $SU(2)_L$ doublet or a
vector-like state -- could be of phenomenological
relevance~\cite{Mirabelli:1999ks,ArkaniHamed:2002qx,Han:2003wu,Agashe:2006at} 
when produced either in pairs~\cite{AguilarSaavedra:2005pv,Contino:2008hi,AguilarSaavedra:2009es} or 
singly~\cite{Han:2003wu,Alwall:2006bx}. For instance, the presence of a fourth generation is not excluded by
precision electroweak and flavor data~\cite{Alwall:2006bx,Kribs:2007nz,Soni:2008bc,Chanowitz:2009mz,Holdom:2009rf,Soni:2009fg}, 
although direct searches at the Tevatron provide some lower bounds on the quark masses. The current 
95\% confidence level limits are~\cite{Amsler:2008zzb},
\begin{equation}
m_{t^\prime} > 256~\mbox{GeV} \, , \qquad m_{b^\prime} > 128~\mbox{GeV} \,
\end{equation}
with this limit on the $b^\prime$ mass corresponding to a charged current
decay. In particular scenarios, where the masses are large and/or there is sizeable mixing
with 3rd generation quarks, electroweak production at the LHC can in
fact be larger than the usual strong pair production.

\begin{figure}[t!]
\centering
\subfigure[]{\includegraphics[scale=.9]{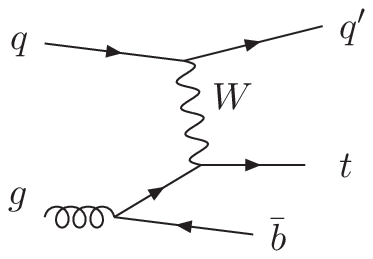}}
\hspace*{1.5cm}
\subfigure[]{\includegraphics[scale=.9]{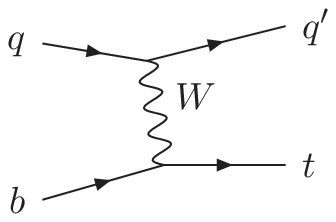}}
\caption{(a) One of two diagrams entering the LO calculation of the
$2 \to 3$ single top process; (b) the single LO diagram for the
$2 \to 2$ process. The NLO calculations presented in this paper correspond
to dressings of these diagrams with additional real and virtual gluon radiation.}
\label{fig:one}
\end{figure}

In a recent letter~\cite{Campbell:2009ss} we have reported on the
first NLO calculation of single-top production in the $t$ channel
based on the $2\to 3$ leading order process, where both heavy quark
masses are kept finite~(Fig.~\ref{fig:one}(a)). This extends previous
NLO (QCD or EW) results which are based on the $2\to 2$ leading order
process~\cite{Bordes:1994ki,Stelzer:1997ns,Harris:2002md,Campbell:2004ch,
Cao:2004ap,Cao:2005pq,Frixione:2005vw,Beccaria:2006ir,Beccaria:2008av}. A
comparison of the two calculations allows a first study of their
consistency across a broad kinematic range.  In this work we perform a
thorough analysis of the cross sections for single-top production, for
both the $2 \to 2$~(Fig.~\ref{fig:one}(b)) and $2 \to 3$ based
calculations, including scale dependence and PDF uncertainties, as
described in Section~\ref{sec:singletop}. In
Section~\ref{sec:smresults} we present results as a function of the
top mass, with values as high as 2 TeV in order to provide predictions
for a $t^\prime$. In Section~\ref{sec:4results} we present for the
first time the corresponding $t$-channel cross sections for the
production of a top quark in association with a $b^\prime$ and for a
4th generation doublet $t^\prime b^\prime$.  In order to illustrate
the sensitivity of this channel, we compare these with the NLO
predictions for strong pair production. Finally, we conclude with a
brief discussion of our results and perspectives for improvement.

\section{Methodology}
\label{sec:singletop}
We perform a study similar to that of Ref.~\cite{Cacciari:2008zb}
for the case of $t \bar t$ production, presenting our
results in the form
\begin{equation}
\sigma=
\sigma_{\rm central} \; {}_{-\Delta{\sigma}_{\mu -}}^{+\Delta{\sigma}_{\mu +}}
~_{-\Delta{\sigma}_{\sss PDF-}}^{+\Delta{\sigma}_{\sss PDF+}}\,,
\end{equation}
where $\sigma_{\rm central}$ is our best prediction and 
$\Delta{\sigma}_{\mu\pm}$ and $\Delta{\sigma}_{\sss PDF\pm}$
quantify the uncertainties due to unknown higher orders
and our incomplete knowledge of the non-perturbative PDFs, respectively.
$\sigma_{\rm central}$ is tabulated/plotted as a function of the heaviest  quark mass, $m_{t^{(\prime)}}$ or $m_{b^{(\prime)}}$. 
We always quote each source of uncertainty explicitly, except for
the plots (Figures~\ref{fig:1}--\ref{fig:cross})
where the total uncertainty band is the direct sum of the two.
Our choice of parameters for the best prediction,
and the prescription that we have used to evaluate the uncertainties,
are specified below.

\subsection{Best prediction}
Our best prediction, $\sigma_{\rm central}$, is computed using a
similar choice of renormalisation and factorisation scales to what was
previously advocated in~\cite{Campbell:2009ss}. For the $2 \to 3$
calculation this corresponds to choosing separate scales on the light
and heavy quark lines ($\mu_l$ and $\mu_h$ respectively), where $\mu_l
= (m_t+m_b)/2$ and $\mu_h=(m_t+m_b)/4$.  The $2 \to 2$ calculation
uses a single overall scale, $m_t/2$, and for both processes the
renormalisation and factorisation scales are equal.  For a heavy quark
with a mass similar to that of the top quark we have shown in
Ref.~\cite{Campbell:2009ss} that this is a judicious choice of scales.
A similar scale choice is also suggested in other processes that
feature initial-state gluon splitting into a heavy quark pair, such as
$gg \to b{\bar b}H$~\cite{Maltoni:2003pn,Campbell:2004pu}.  For much larger top quark
masses we find similar conclusions, although even smaller scales for
the heavy quark line might be preferred. However, for the sake of
simplicity, we maintain the same choice since for very high quark
masses the accuracy of our calculations will in any case be affected
by potentially important threshold resummation
effects~\cite{Kidonakis:2006bu,Kidonakis:2007ej}.

We present results for two choices of PDF family,
CTEQ6.6~\cite{Nadolsky:2008zw} and MSTW2008~\cite{Martin:2009iq}.  Our
central prediction for each of these choices corresponds to the best
fit for each family.   It is important to note that CTEQ and MSTW
use different $b$ masses in their fits ($m_b=4.5$~GeV and $m_b=4.75$~GeV
respectively). For the results presented here we have used the value
prescribed by the PDF set for their respective predictions. In
Sec.~\ref{ssec:massdep} the uncertainty coming from the heavy quark masses  is
addressed.  We note that, since these are 5-flavour PDF sets, we pass
to the 4-flavor scheme necessary for a consistent calculation of the
$tb$ and $t^\prime b$ processes by including the counterterms of
Ref.~\cite{Cacciari:1998it}.  For $t b^\prime$ and $t^\prime b^\prime$
we always use the pure 5-flavor PDFs.

Though mixings from the CKM matrix are fully included in our calculation, 
for simplicity we present results with the $V_{ij}$ relative to the  third and possibly the 
fourth generation set to unity, 
\ie,  $V_{tb} = V_{tb^\prime} = V_{t^\prime b} = V_{t^\prime b^\prime} = 1$ is assumed.
 
\subsection{Uncertainty from higher orders}
The uncertainty from uncalculated higher orders in the perturbative expansion
is estimated by varying the factorisation and the renormalisation scales
$\muF$ and $\muR$ independently around the central scale choice, $\mu_0$.
As discussed above, in the $2 \to 3$ calculation we have
$\mu_0 = \mu_l$ for the light line and $\mu_0 = \mu_h$ for the heavy line,
whereas the $2 \to 2$ calculation uses $\mu_0 = m_t/2$. 

The values of $\mu_F$ and $\mu_R$ that we range over are specified by,
\begin{equation}
(\mu_F, \mu_R) \in \Bigl\{ (2\mu_0, 2\mu_0), \, (2\mu_0, \mu_0), \,
 (\mu_0, 2\mu_0), \, (\mu_0, \mu_0/2), \, (\mu_0/2, \mu_0), \,
 (\mu_0/2, \mu_0/2) \Bigr\} \;.
\end{equation}
In this way we ensure that none of the ratios, $\mu_F/\mu_0$, $\mu_R/\mu_0$ and
$\mu_F/\mu_R$, is outside the interval $[\frac{1}{2}, 2]$. These ratios naturally appear
as arguments of logarithms at NLO, so restricting them in this way is motivated
by the requirement of good perturbative behaviour.

The uncertainties are then defined with respect to this set of variations as,
\begin{align}
\Delta{\sigma}_{\mu +} = \phantom{-}&\max_{\{\mu_F,\mu_R\}}
\Big[\sigma(\mu_F,\mu_R)-\sigma_{\rm central}\Big]\,,
\label{eq:muplus}
\\
\Delta{\sigma}_{\mu -} = -&\min_{\{\mu_F,\mu_R\}}
\Big[\sigma(\mu_F,\mu_R)-\sigma_{\rm central}\Big]\,.
\label{eq:muminus}
\end{align}

\subsection{PDF uncertainty}
We estimate the PDF uncertainty on our predictions by making use of the
additional sets that are supplied by the CTEQ and MSTW collaborations
for that purpose. There are 44 such sets for the CTEQ6.6~\cite{Nadolsky:2008zw}
PDFs and 30 for the MSTW~\cite{Martin:2009iq} ones, that are organized in
pairs (which we call here $set_{+i}$ and $set_{-i}$).
We follow the prescription of
MSTW~\cite{Martin:2009iq} and determine asymmetric uncertainties in
the form,
\begin{eqnarray}
&&\Delta{\sigma}_{\sss PDF+} =
\sqrt{\sum_i\Big(\max\Big[{\sigma}(set_{+i}) - {\sigma}(set_0),
{\sigma}(set_{-i}) - {\sigma}(set_0), 0\Big]\Big)^2} \,,
\label{eq:PDFplus}
\\
&&\Delta{\sigma}_{\sss PDF-} =
\sqrt{\sum_i\Big(\max\Big[{\sigma}(set_0) - {\sigma}(set_{+i}),
{\sigma}(set_0) - {\sigma}(set_{-i}), 0\Big]\Big)^2} \,.
\label{eq:PDFminus}
\end{eqnarray}
where all cross sections are evaluated using our central scale choices
and the usual (best fit) PDF set is labelled by $set_0$. This procedure
yields an uncertainty at approximately the 90\% confidence level.

\section{Single top and $t^\prime$ results}
\label{sec:smresults}

We present here the cross sections for single $t$ and $t^\prime$
production, obtained from both the $2\to 2$ and $2\to3$ calculations,
as implemented in the Monte Carlo program
MCFM~\cite{Campbell:1999ah,Campbell:2004ch,Campbell:2009ss}. We show
results for Run II of the Tevatron and for two LHC energies, 10 TeV
and 14 TeV.  For single-top production we show results for a number of
different values of the top mass around the current best
determination~\cite{:2009ec} and for $t^\prime$ production we
investigate quark masses as large as 2 TeV (at the LHC). The 
dependence on the bottom mass is addressed in a  subsection dedicated
to the best SM predictions, Sec.~\ref{ssec:massdep}. As mentioned earlier, we remind the reader that in order 
to make a fair comparison between the $2\to 2$ and $2\to3$ calculations, the bottom mass has to be consistent
with the value assumed in the PDFs, \ie, $m_b=4.5 (4.75)$ GeV for the CTEQ6.6 (MSTW2008) PDFs.
Here and in the rest of this paper, by `quark masses' we mean the pole masses that
are the input in the modified $\overline{\textrm{MS}}$ renormalisation
scheme that we adopted in the calculation of the virtual contributions
at NLO. The cross sections and uncertainties are collected in
Appendix~\ref{app:A} in Tables~\ref{tab:1} and~\ref{tab:2} (Tevatron),
Tables~\ref{tab:3} and~\ref{tab:4} (LHC, 10 TeV) and
Tables~\ref{tab:5} and~\ref{tab:6} (LHC, 14 TeV).  For each machine,
the two tables correspond to our two choices of central PDF.  Note
that, at the LHC, the rates for production of a heavy top (or
$t^\prime$) are different from those of its antiparticle and the two
are thus studied separately. The CTEQ6.6 results for the sum of top
(or $t^\prime$) and anti-top (or $\bar t^\prime$) production are
illustrated in Figures~\ref{fig:1} (Tevatron), \ref{fig:3} (LHC, 10
TeV) and~\ref{fig:5} (LHC, 14 TeV), where we also show the NLO rates
for the corresponding strong pair production for comparison.

A number of global features are evident from these results.  Firstly,
the central cross sections predicted by the $2 \to 2$ and $2 \to 3$
processes differ by 5\% or less, both at the Tevatron and at the LHC,
for masses around the top quark.  At the Tevatron, the difference is
well within the combined uncertainty from higher orders and PDFs, so
we conclude that the two calculations are consistent.  At the LHC (10
and 14 TeV) the consistency is marginal, in particular because the
uncertainties from the PDFs are (almost) 100\% correlated between the
two approximations.  We stress therefore that in a combined estimate of
the total production cross section, the PDF errors should \emph{not}
be summed.

For larger masses, \ie~for $t^\prime$ production, the differences are
much larger. For a mass of 1 TeV, the $2 \to 2$ prediction using the
CTEQ6.6 PDF set is almost twice as large at the Tevatron and $20\%$
bigger at the LHC.  However for such large top masses it could well be
that the logarithm that is implicitly resummed in the bottom quark
distribution function might become relevant or that an even smaller
scale choice should be used.  Nevertheless we see that the differences
between the two calculations can still be accounted for by their
uncertainties (with the same PDF caveats as above) and at this stage
it is difficult to establish which of the two calculations, the $2 \to
2$ or the $2 \to 3$ one, might be more accurate for the total cross section.

\FIGURE[!t]{
  \epsfig{file=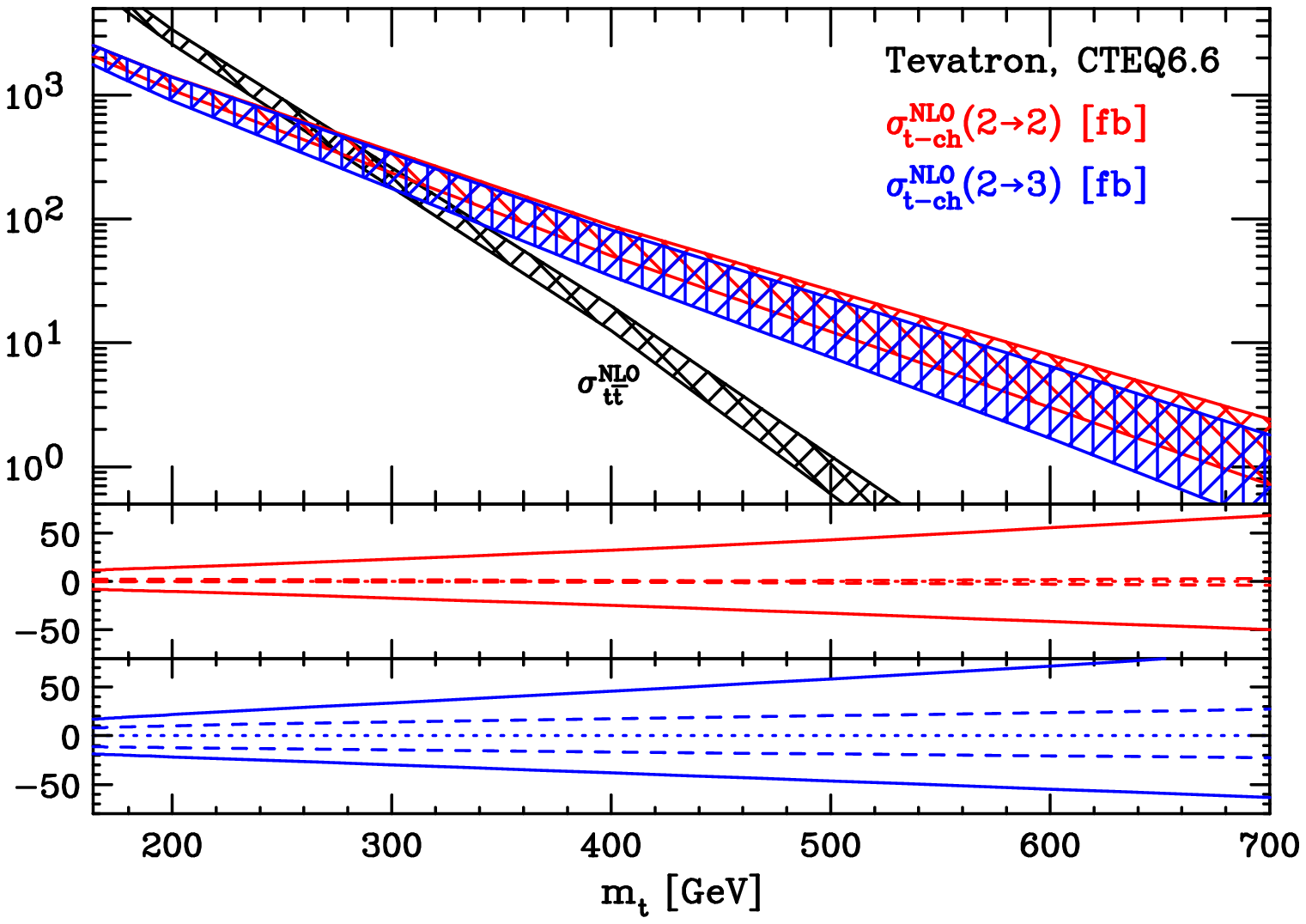, width=0.70\textwidth}
  \caption{ Cross sections (fb) at the Tevatron Run II for the sum of
    top and anti-top quark production in the $t$ channel, as a
    function of the top mass obtained with the CTEQ6.6 PDF set and
    $V_{t^{(\prime)}b}=1$ in the $2\to2$ and $2\to3$ schemes.  Bands
    are the total uncertainty (scale+PDF). In the lower plots, dashed
    is scale uncertainty, solid is scale + PDF.  The corresponding
    data is collected in Table~\protect\ref{tab:1}.  }
  \label{fig:1}
}

\FIGURE[!t]{
  \epsfig{file=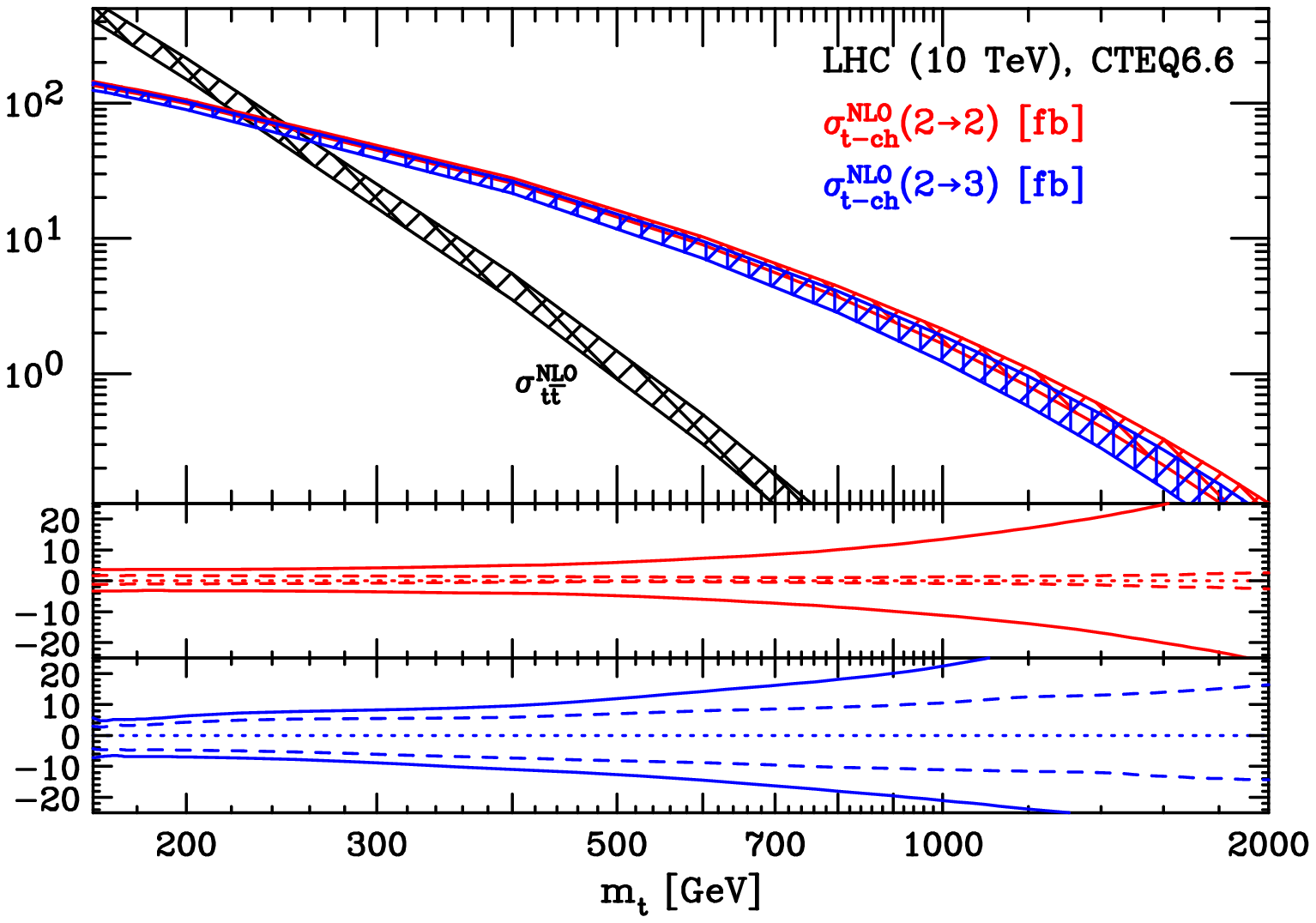, width=0.70\textwidth}
  \caption{
The same as Figure~\protect\ref{fig:1} but for the LHC 10 TeV.
The corresponding data is collected in Table~\protect\ref{tab:3}. 
\label{fig:3}}
}

\FIGURE[!t]{
  \epsfig{file=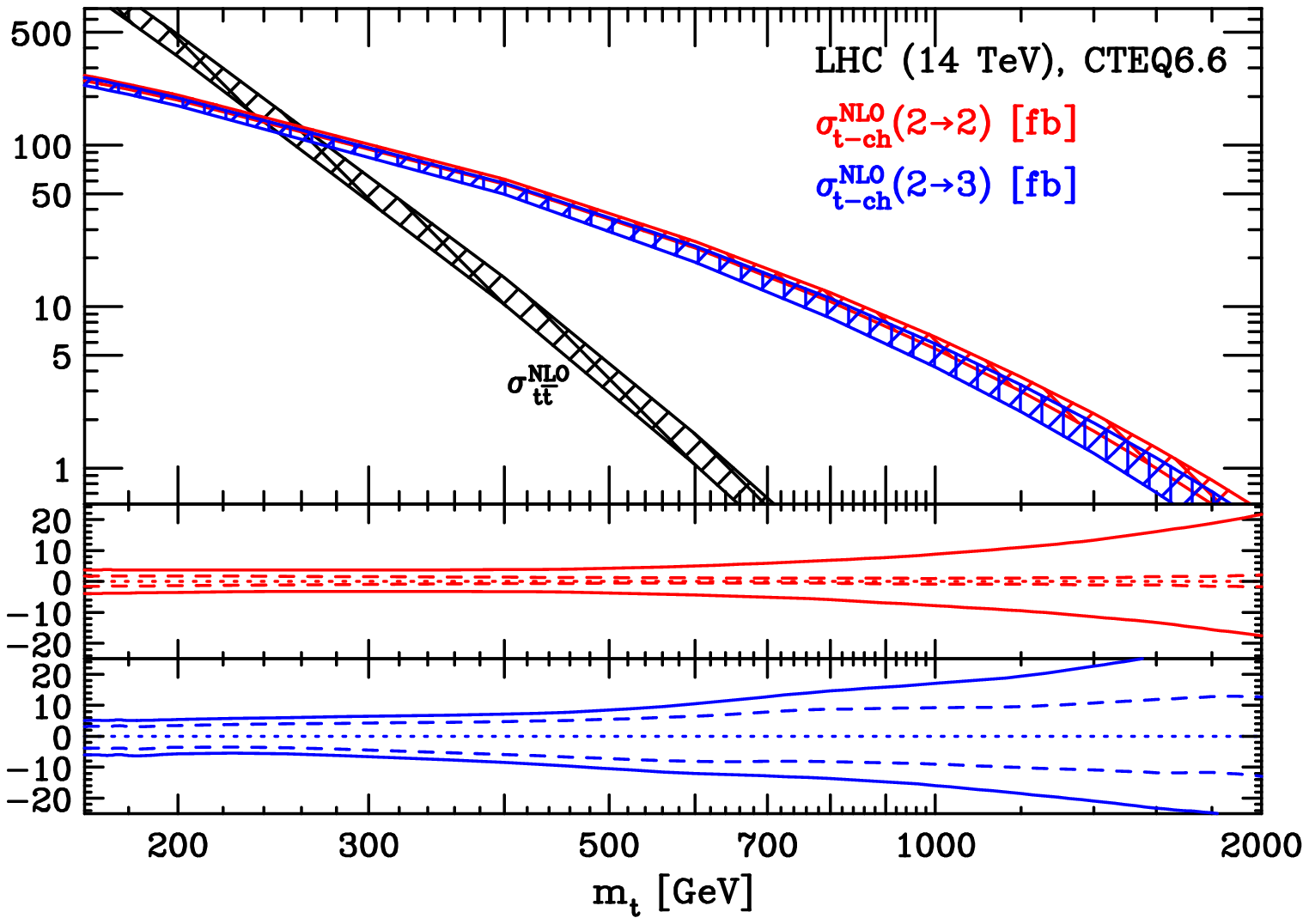, width=0.70\textwidth}
  \caption{ 
The same as Figure~\protect\ref{fig:1} but for the LHC 14 TeV.
The corresponding data is collected in  Table~\protect\ref{tab:5}. 
\label{fig:5}}
}

The uncertainty coming from higher orders is estimated to be much larger for the
$2 \to 3$ process, particularly at the Tevatron. This is somewhat expected, since the perturbative
series for this process begins at ${\cal O}(\alpha_s)$ rather than the purely electroweak leading
order of the $2 \to 2$ calculation. Our estimate ascribes uncertainties that are typically at
the level of a few percent, except for the $2 \to 3$ calculation at the Tevatron where they
range from about $10\%$ for the top quark to as large as $30\%$ for a 1 TeV $t^\prime$.

Results from the two different PDF sets are generally in good agreement with one another
at NLO. For the top quark, differences are at the level of a couple of percent or less in the
$2 \to 2$ calculation, a few percent for the $2 \to 3$ process and are smaller at the LHC than
at the Tevatron. 

The uncertainty on the cross sections deriving from the PDFs is different between the
Tevatron and the LHC. At the Tevatron the PDF uncertainty is particularly large, with the
CTEQ6.6 set yielding approximately a $10\%$ uncertainty for the top quark (slightly smaller
for MSTW), whilst for a $t^\prime$ it can be considerably larger. This simply represents the limitations
of current global PDF determinations, where the gluon distribution at large momentum fractions
is not very well constrained by data. This explains
not only the much larger percentage uncertainties, but also the greater difference between the
CTEQ6.6 and MSTW predictions. For such large masses the cross sections presented here clearly have
little phenomenological interest at the Tevatron.

At the LHC the PDF uncertainty on the $2 \to 2$ single top quark cross section is comparable to that 
coming from the unknown higher orders, whilst it is somewhat smaller for the $2 \to 3$ calculation.
For single $t^\prime$ production the two sources of uncertainty are comparable in the $2 \to 3$ case, but
the uncertainty from the PDFs clearly dominates for the $2 \to 2$ process.

\subsection{Single top cross sections in the SM:  bottom and top mass dependence}
\label{ssec:massdep}

For the case of single top production in the Standard Model, whose cross section can be
predicted quite accurately, it is important to investigate in detail its dependence on both
heavy quark masses. In the following we discuss the bottom and top mass effects independently,
having checked explicitly that this is a very good approximation.

To our knowledge the bottom quark mass dependence of the total single top
cross section has never been addressed in detail. There are two different ways in which
the bottom mass can enter the final results, \ie, through  logarithmic and power correction terms.
In the $2 \to 3$ calculation both effects explicitly depend on the pole mass parameter that
is already present at LO in the matrix element (and in  the phase space boundaries). In the
$2 \to 3$ based computation it is therefore trivial to quantify the bottom mass dependence.
On the other hand, disentangling the relative impact of the logarithmic terms from those that
are power suppressed requires some analytical work.
In the $2 \to 2$ calculation the situation is reversed. The two sources are separated from 
the start: the effect of the logarithms is resummed in the bottom PDF while the power-like terms
at NLO come from the (subtracted) real correction diagrams $q g \to t b q'$.  The effect of the
$b$ mass from the latter source has already been studied, see for instance
Ref.~\cite{Sullivan:2004ie}, and found to be very small. We have checked that changing the
bottom mass from $\sim$5~GeV to zero results in a difference below 0.5\%.
However the logarithmic corrections, which have so far been neglected, are quite sizeable.   
In the $2 \to 2$ calculation the logarithmic dependence on the bottom quark mass is
``hidden" in the starting condition for the evolution of the $b$-PDF. To study its  impact we have
generated various sets of PDFs for different bottom masses in the range 4 GeV$<m_b<$5 GeV,
using the evolution code provided by the CTEQ collaboration. As a result we find that 
the $ 2 \to 2$ cross sections on average decrease by about 1.2\%, 0.86\%, 0.80\% per 100 MeV increase of $m_b$ at the 
Tevatron, LHC at 10 TeV and 14 TeV, respectively. Such a dependence is fully reproduced 
by the $2 \to 3$ calculation which gives similar (corresponding) results, namely 1.0\%, 0.83\% and 0.76\%. 
We conclude that the $b$ mass dependence should be included as a source of uncertainty in the 
final predictions for the SM total cross sections.

As far as the top quark mass dependence is concerned, this can be studied easily in both the
$2 \to 2$ and $2 \to 3$ calculations. 
Here we provide formulae that can be used to obtain  the cross section and the corresponding uncertainties 
for any top quark mass in the range 164~GeV~$<m_t<180$~GeV,  to an accuracy better than 1\%.
We fit the mass dependence of the cross section using a quadric centered on 
$\sigma_0=\sigma(m_t=172\textrm{ GeV})$,
\begin{equation}
  \sigma(m_t)=\sigma_0\Bigg[1+\bigg(\frac{m_t-172}{172}\bigg)A
   +\bigg(\frac{m_t-172}{172}\bigg)^{\!\!2}B\Bigg] \;,
\end{equation}
where $m_t$ is measured in GeV.
The (dimensionless) coefficients $A$ and $B$ for both the 
$2 \to 2$ and $2 \to 3$ processes are given in
Table~\ref{tab:AB}.

\begin{small}
\renewcommand{\arraystretch}{1.1}
\TABULAR[th]{ccccccccccc}
{
\toprule[0.08em]
\multicolumn{2}{c}{\multirow{2}{*}{Collider}}&\multirow{2}{*}{Born}&&\multicolumn{3}{c}{MSTW2008}&&\multicolumn{3}{c}{CTEQ6.6}\\
&&&&$\sigma_0$&$A$&$B$&&$\sigma_0$&$A$&$B$\\
\midrule[0.05em]
\multirow{2}{*}{Tevatron    }&\multirow{2}{*}{$t=\bar{t}$}&$2\to 2$ &&$   994  ^{+    24 }_{-     2}~^{+    61 }_{-    52}$&-3.01 &   5.2&&$   981  ^{+    23 }_{-     3}~^{+    98 }_{-    82}$& -2.97 &   5.0\\
&&$2\to 3$ &&$   942  ^{+    86 }_{-   113}~^{+    53 }_{-    43 }$& -3.11 &   4.9 &&$   935  ^{+    82 }_{-   107}~^{+    88 }_{-    74 }$& -3.08 &   5.3\\
\midrule[0.05em]
\multirow{4}{*}{LHC (10 TeV)}&\multirow{2}{*}{$t$        }&$2\to 2$ &&$ 84.4  ^{+  1.4 }_{-  1.0}~^{+  1.1 }_{-  1.0}$& -1.48 &   1.1 &&$ 83.5  ^{+  1.4 }_{-  1.1}~^{+  1.5 }_{-  1.7}$& -1.50 &   1.0\\
\vspace{2pt}
&&$2\to 3$ &&$ 80.3  ^{+  3.2 }_{-  3.7}~^{+  1.1 }_{-  1.0 }$& -1.60 &   2.2 &&$ 79.8  ^{+  2.9 }_{-  3.4}~^{+  1.4 }_{-  1.6 }$& -1.54 &   3.3\\
&\multirow{2}{*}{$\bar{t}$  }&$2\to 2$ &&$ 48.3  ^{+  0.8 }_{-  0.5}~^{+  0.7 }_{-  1.0}$& -1.57 &   1.7 &&$ 46.6  ^{+  0.8 }_{-  0.5}~^{+  1.0 }_{-  1.1}$& -1.58 &   1.2\\
&&$2\to 3$ &&$ 45.4  ^{+  1.7 }_{-  2.1}~^{+  0.7 }_{-  1.0 }$& -1.54 &   0.4 &&$ 44.2  ^{+  1.2 }_{-  2.0}~^{+  1.0 }_{-  1.1 }$& -1.68 &   0.9\\
\midrule[0.05em]
\multirow{4}{*}{LHC (14 TeV)}&\multirow{2}{*}{$t$        }&$2\to 2$ &&$154.3  ^{+  2.9 }_{-  2.5}~^{+  2.2 }_{-  2.2}$& -1.32 &   1.1 &&$152.9  ^{+  3.0 }_{-  2.3}~^{+  3.0 }_{-  3.4}$& -1.35 &   2.2\\
\vspace{2pt}
&&$2\to 3$ &&$146.8  ^{+  4.3 }_{-  5.0}~^{+  2.2 }_{-  2.1 }$& -1.40 &   4.6 &&$147.0  ^{+  5.0 }_{-  5.7}~^{+  2.7 }_{-  3.1 }$& -1.35 &   0.9\\
&\multirow{2}{*}{$\bar{t}$  }&$2\to 2$ &&$ 94.2  ^{+  1.6 }_{-  1.5}~^{+  1.2 }_{-  1.9}$& -1.42 &   1.1 &&$ 91.1  ^{+  1.5 }_{-  1.5}~^{+  1.8 }_{-  2.1}$& -1.42 &   0.7\\
&&$2\to 3$ &&$ 88.7  ^{+  2.6 }_{-  3.0}~^{+  1.3 }_{-  1.8 }$& -1.49 &   3.1 &&$ 86.8  ^{+  2.4 }_{-  3.7}~^{+  1.8 }_{-  2.0 }$& -1.46 &   0.9\\

\bottomrule[0.08em]
 }
{
\label{tab:AB}
Parameters to interpolate the cross section (in fb for the Tevatron, pb for the LHC) and the 
corresponding uncertainties for top quark masses around the default value of $\sigma_0=\sigma(m_t=172 \textrm{ GeV})$.
}\
\end{small}

\FIGURE[!t]{
  \epsfig{file=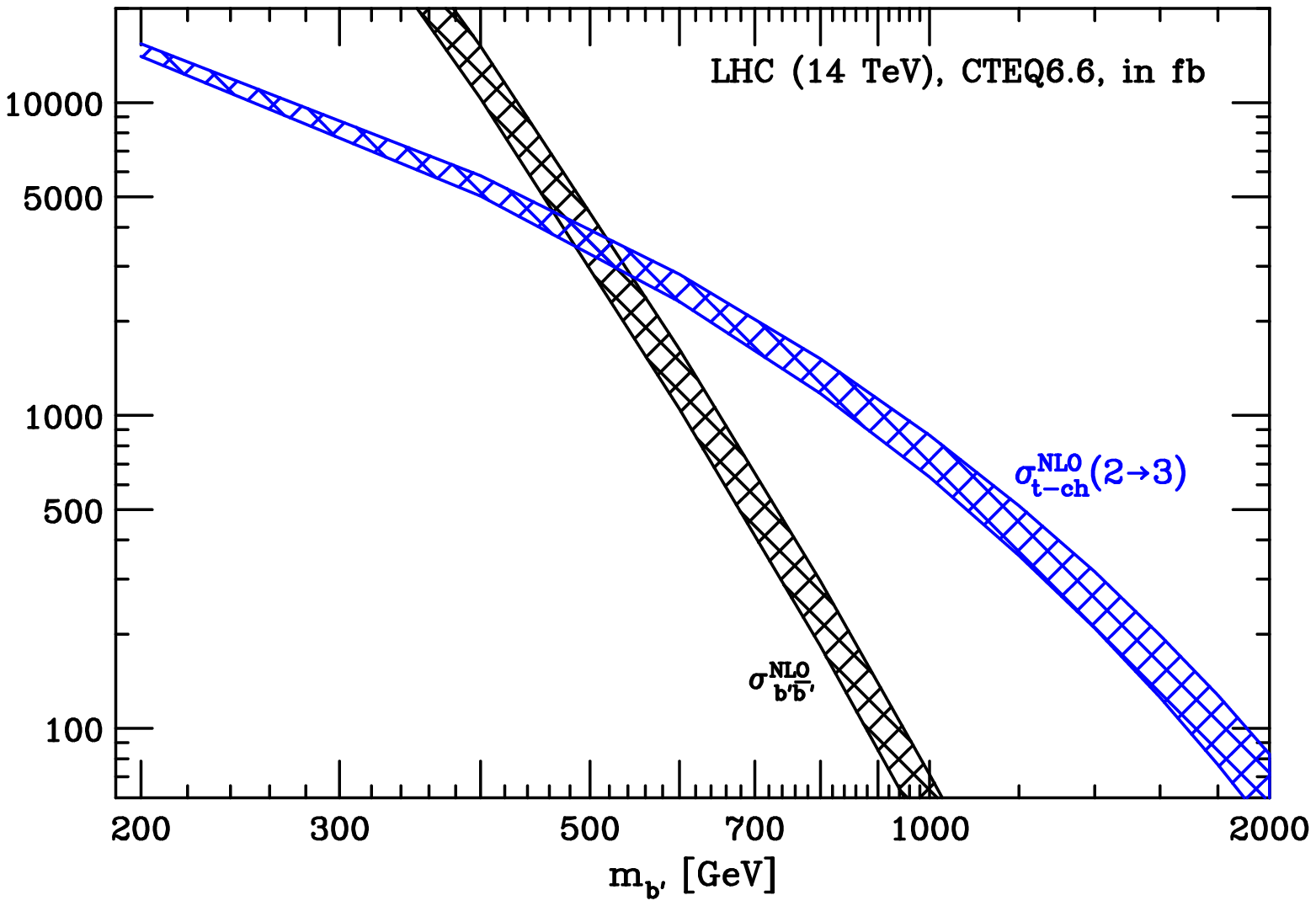, width=0.70\textwidth}
  \caption{Cross sections (fb) at the LHC 14 TeV for $b^\prime \bar t$ plus  $\bar b^\prime t$ production as a function of $m_{b^\prime}$ 
obtained with the CTEQ6.6 PDF set and $V_{tb^{\prime}}=1$. Bands are the total uncertainty (scale+PDF).The corresponding data is collected in Table~\protect\ref{tab:h}. 
}
  \label{fig:h}
}

\FIGURE[!t]{
  \epsfig{file=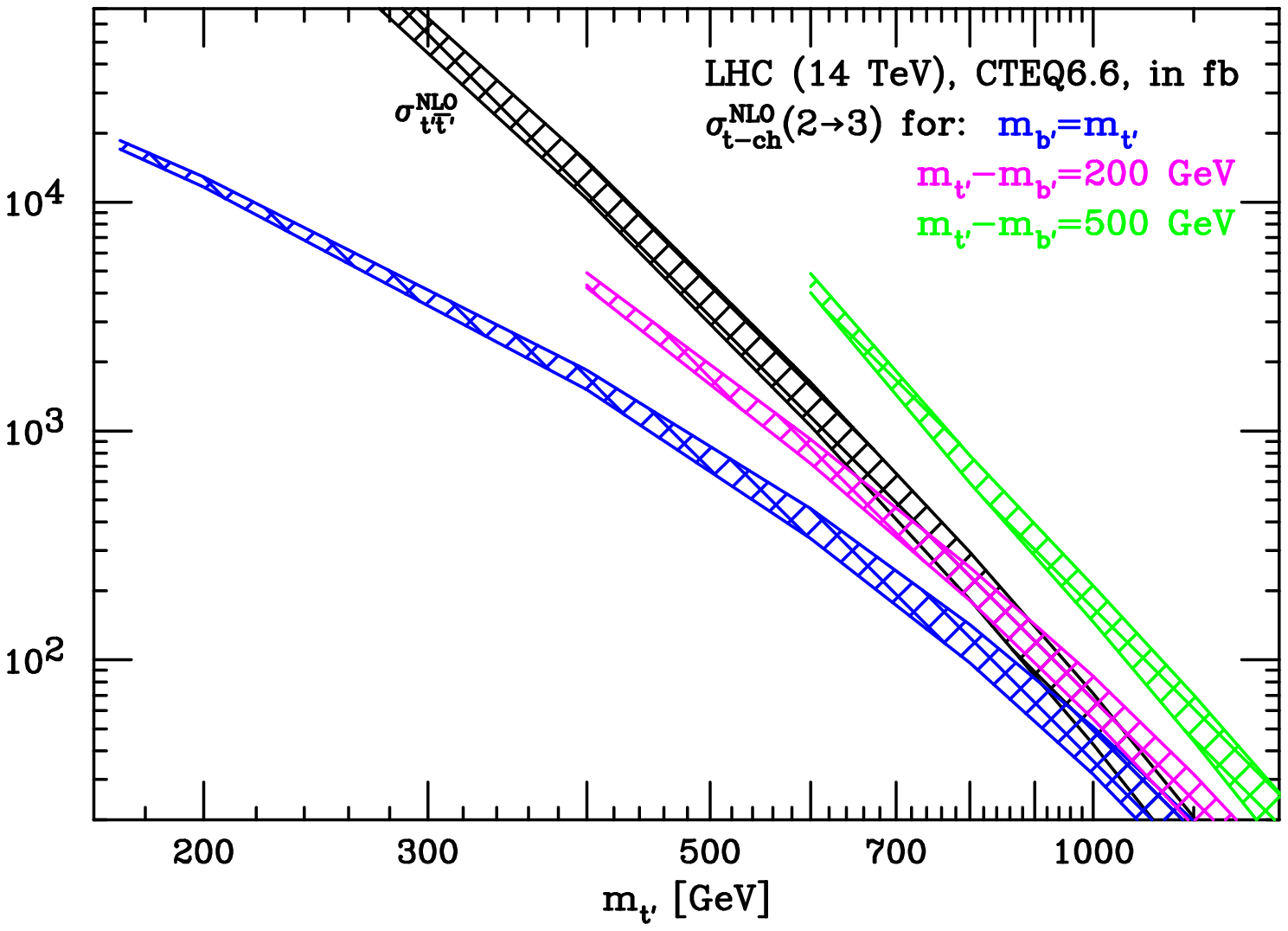, width=0.70\textwidth}
  \caption{ 
   Cross sections (fb) at the LHC 14 TeV for $t^\prime \bar b^\prime$ plus  $\bar t^\prime b^\prime$ production as a function of $m_{t\prime}$ 
obtained with the CTEQ6.6 PDF set and $V_{t^{\prime}b^{\prime}}=1$.  Bands are the total uncertainty (scale+PDF). The corresponding data is collected in Table~\protect\ref{tab:7}. 
}
  \label{fig:7}
}

\FIGURE[!t]{
  \epsfig{file=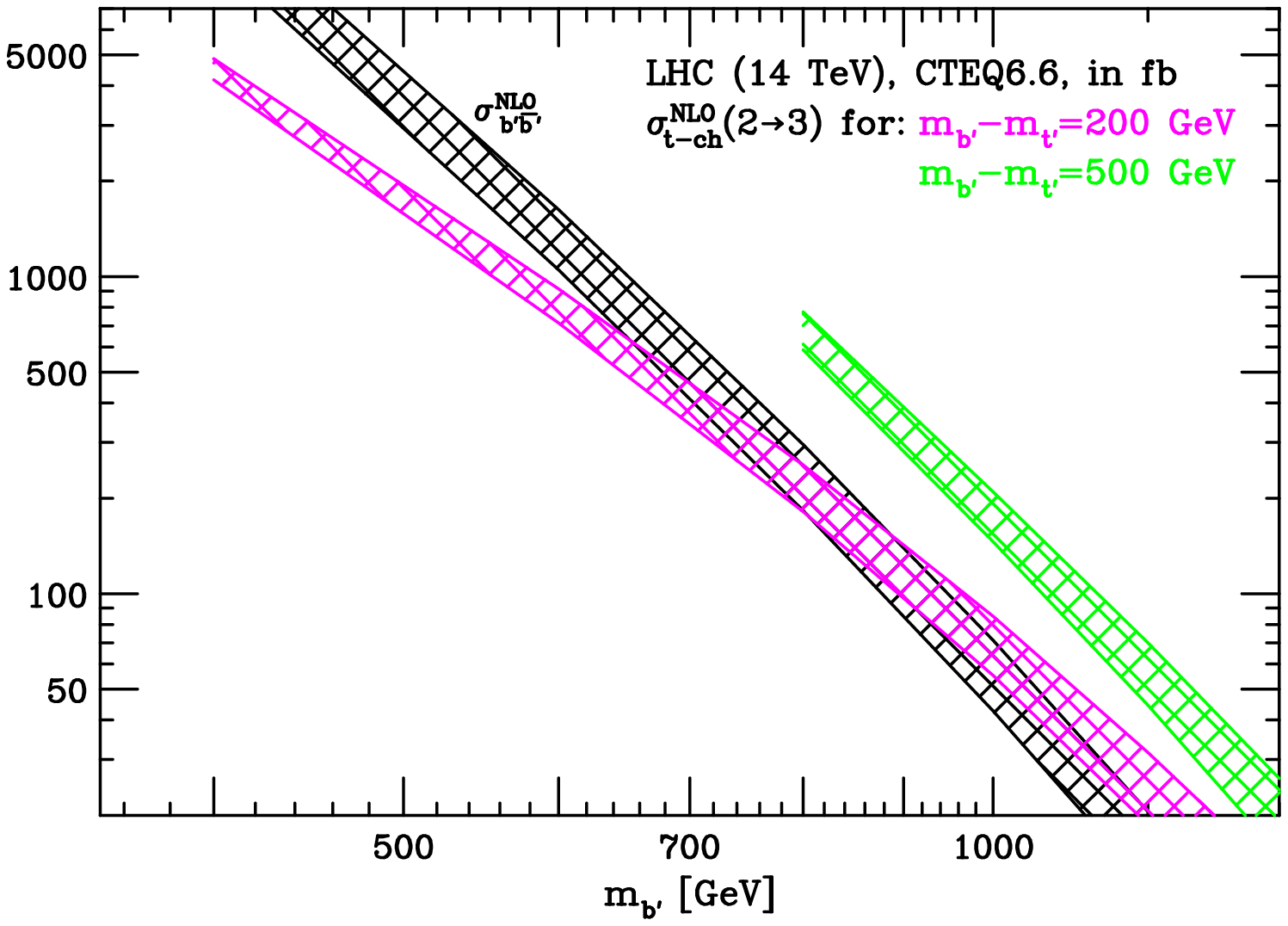, width=0.70\textwidth}
  \caption{Cross sections (fb) at the LHC 14 TeV for $b^\prime \bar t^\prime$ plus  $\bar b^\prime t^\prime$ production as a function of $m_{b^\prime}$ 
obtained with the CTEQ6.6 PDF set and $V_{t^{\prime}b^{\prime}}=1$. Bands are the total uncertainty (scale+PDF). The corresponding data is collected in able~\protect\ref{tab:8}. 
 }
  \label{fig:8}
}

\section{$t  b^\prime$ and $ t^\prime b^\prime$ results}
\label{sec:4results}

Production of a $b^\prime$, together with either a top quark or an
additional $t^\prime$, can be computed at NLO accuracy starting from
the $2 \to 3$ Standard Model single-top process. Since the final state
contains two heavy particles, the expected rates are probably out of
the reach of the sensitivity of the Tevatron and early LHC running.
Therefore we present results for the 14 TeV LHC energy only, but
explore a range of different $t^\prime$ and $b^\prime$ masses.

For $b^\prime \bar{t}$ (and $\bar{b}^\prime t$) production the cross
sections and their uncertainties are tabulated for both choices of PDF
set in Tables~\ref{tab:h} and~\ref{tab:i}, collected in
Appendix~\ref{app:B}. The two sources of uncertainty considered here
are of a comparable size, leading to an overall accuracy running from
a few percent for $m_{b^\prime}=200$~GeV up to about 30\% for the
highest masses considered, once again due to the PDF uncertainty in
that kinematic region. The cross sections for the sum of $b^\prime
\bar{t}$ and $\bar{b}^\prime t$ production including uncertainties are
also plotted in Figure~\ref{fig:h} for the CTEQ6.6 PDF set. For
comparison strong production of $b^\prime$ pairs is also shown.

In the case of $t^\prime b^\prime$ production we consider five
different scenarios: $m_{b^\prime} = m_{t^\prime}$, $m_{t^\prime} -
m_{b^\prime} = 200$~GeV and $m_{t^\prime} - m_{b^\prime} = 500$~GeV as
a function of the $t^\prime$ mass and $m_{b^\prime} - m_{t^\prime} =
200$~GeV and $m_{b^\prime} - m_{t^\prime} = 500$~GeV as a function of
the $b^\prime$ mass.  Note that mass splittings of this magnitude in a
fourth generation would induce a shift in the $\rho$ parameter
incompatible with precision EW measurements~\cite{:2005ema}. However,
as our purpose is to provide benchmark cross-sections useful for
constructing new models we neglect such constraints here. For
instance, introducing extra matter could compensate for these effects
and/or the $b'$ and $t'$ might be higher representations of the
$SU(2)_L$ symmetry group (with the only constraint being the one unit
of charge between two quarks).

The cross sections and uncertainties in the five
scenarios are also tabulated in the appendices, for the first three
cases together in Tables~\ref{tab:7} and~\ref{tab:f} and for the
latter two together in Tables~\ref{tab:8} and~\ref{tab:g}, using the
CTEQ6.6 and MSTW2008 PDF sets respectively.  The CTEQ6.6 results for the
first three scenarios are illustrated in Figure~\ref{fig:7} as the sum
of $t^\prime \bar b^\prime$ and $\bar t^\prime b^\prime$ production,
where we also show the NLO rates for $t^\prime \bar t^\prime$ for
comparison. The latter two scenarios are plotted in Figure~\ref{fig:8}
together with the NLO $b^\prime\bar b^\prime$ cross sections and
uncertainties for comparison.

We conclude this section by briefly commenting on the symmetry
properties of the results for the cross sections.  $\mathcal{CP}$
invariance implies that simultaneously interchanging the $t'$ and $b'$
masses, together with the chirality of the $Wtb$ vertex from left- to
right-handed, gives the same cross section. By performing either a
$\mathcal{C}$ or $\mathcal{P}$ transformation individually, one can
pass from the case with $m_{t^\prime} - m_{b^\prime} = 200 (500)$~GeV
to the case with $m_{b^\prime} - m_{t^\prime} = 200 (500)$~GeV, and
vice-versa.  It is interesting to note that the above cases, while not
related by a symmetry, lead to similar total rates. The differences
arise from angular dependences in the matrix elements that are
proportional to $\beta$, the velocity of the heavy quarks. Since these
terms are integrated over the available phase space and they are
themselves small for high heavy quark masses, they result in only
minor differences.

\section{Conclusions}
\label{sec:conclusions}

\begin{figure}[t!]
\centering
    \epsfig{file=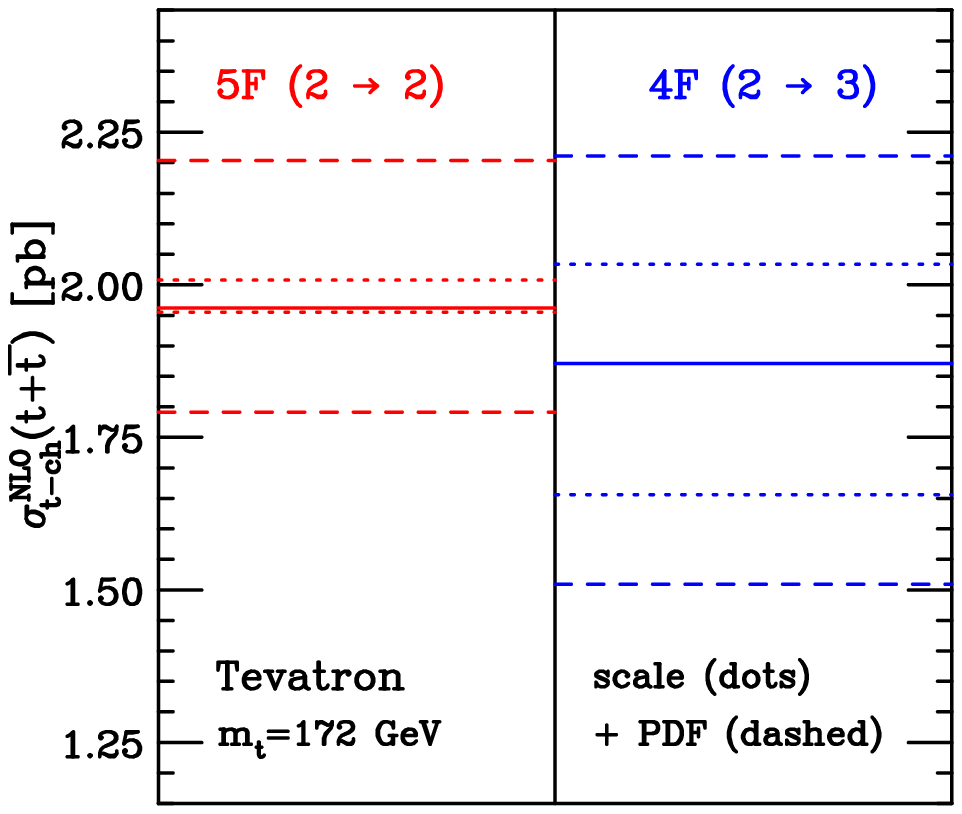, width=0.32\textwidth}
    \epsfig{file=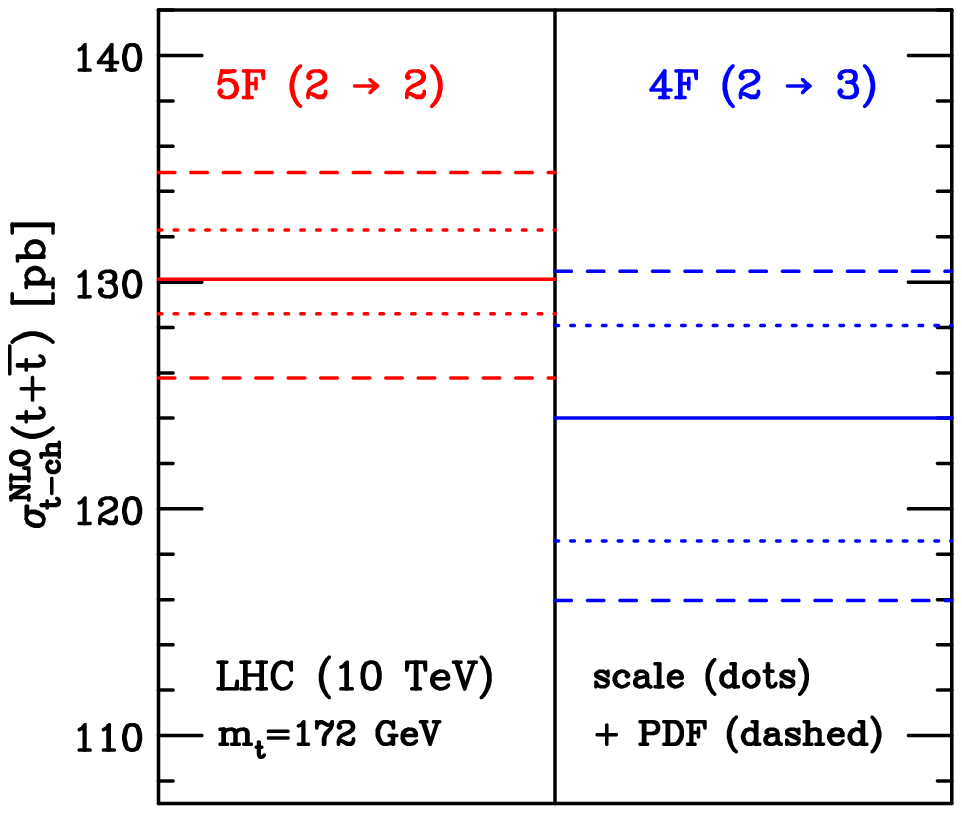, width=0.32\textwidth}
    \epsfig{file=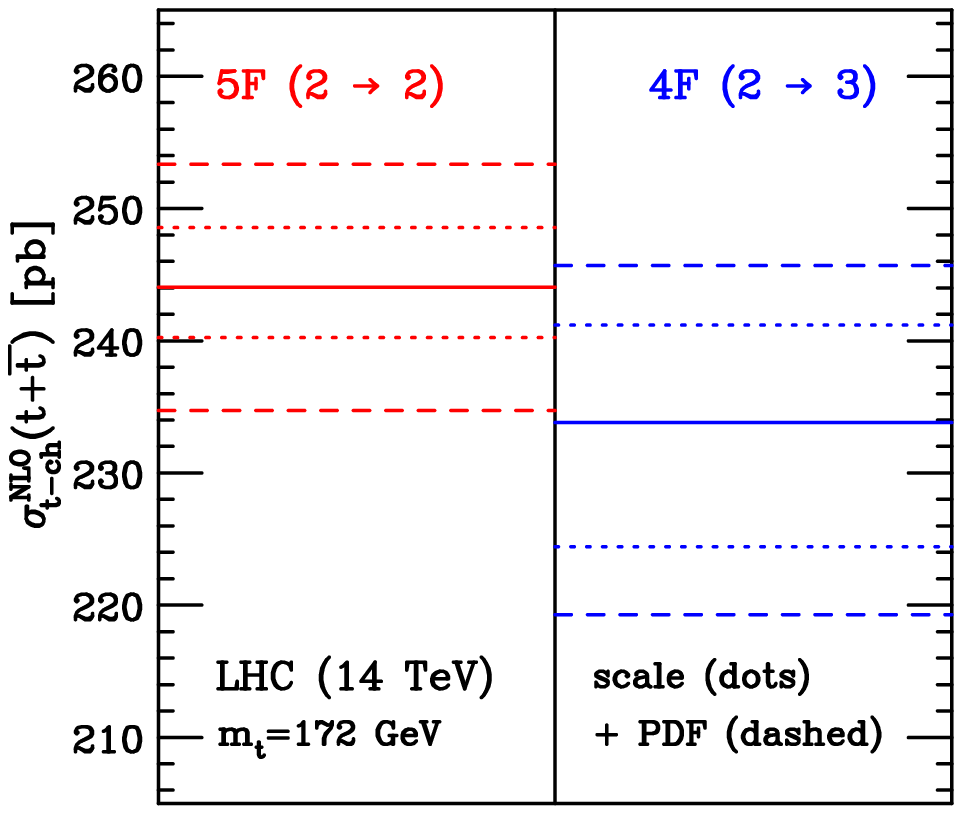, width=0.32\textwidth}
    \caption{Total cross section at NLO for the $2\to2$ and $2\to3$ at
      the Tevatron (top), LHC 10 TeV (bottom-left) and LHC 14 TeV
      (bottom-right).}
  \label{fig:cross}
\end{figure}

The recent discovery of single top production at the Tevatron opens the door to more extensive
studies of this final state both there and at the LHC. In this paper we have presented
an up-to-date and systematic study of both the cross sections that should be expected in this
channel and their associated theoretical uncertainties. Cross sections have been computed at NLO
accuracy in the strong coupling, starting from two Born approximations corresponding to $2 \to 2$
and $2 \to 3$ scattering processes.
Our best predictions for $t$-channel single top cross sections in the $2 \to 2$ and $2 \to 3$ schemes,
with $m_t=172 \pm 1.7$~GeV, $m_b=4.5 \pm 0.2$~GeV and computed using the CTEQ6.6 PDF set, are:
\begin{center}
\begin{tabular}[h]{r@{\qquad}r@{}l@{}l@{}l@{}l@{\qquad}r@{}l@{}l@{}l@{}l}
\toprule[0.08em]

$\sigma_{\rm t-ch}^{\textrm{NLO}}(t+\bar t)$&\multicolumn{5}{c}{$2 \to 2$ (pb)$\quad$} & \multicolumn{5}{c}{$2 \to 3$ (pb)$\quad$}\\
\midrule[0.05em]
\vspace{2pt}
Tevatron Run II & $1.96$&$~^{+0.05 }_{-0.01}$&$~^{+0.20 }_{-0.16}$ &$~^{+0.06}_{-0.06}$&$~^{+0.05}_{-0.05}$& $1.87$&$~^{+0.16 }_{-0.21}$&$~^{+0.18 }_{-0.15}$&$~^{+0.06}_{-0.06}$&$~^{+0.04}_{-0.04}$  \\
\vspace{2pt}
LHC (10 TeV)    & $130$&$~^{+2 }_{-2}$&$~^{+3 }_{-3}$&$~^{+2}_{-2}$&$~^{+2}_{-2}$& $124$&$~^{+4}_{-5}$&$~^{+2 }_{-3}$&$~^{+2}_{-2}$&$~^{+2}_{-2}$ \\
LHC (14 TeV)    & $244$&$~^{+5 }_{-4}$&$~^{+5 }_{-6}$ &$~^{+3}_{-3}$&$~^{+4}_{-4}$& $234$&$~^{+7 }_{-9}$&$~^{+5 }_{-5}$&$~^{+3}_{-3}$&$~^{+4}_{-4}$ \\
\bottomrule[0.08em]
\end{tabular}
\end{center}
The first two uncertainties are computed according to the procedure outlined in
Section~\ref{sec:singletop} and we have used CTEQ6.6 in order to
provide the most conservative predictions. These results are also depicted in the 
plots of Figure~\ref{fig:cross}. The third and fourth uncertainties are related to the top mass and bottom
mass uncertainties, respectively. 

As the results in the two schemes are in substantial agreement and a priori provide equally
accurate though different theoretical descriptions of the same
process, one could try to combine them. We think that this is a
legitimate approach (once correlations among the theoretical errors,
scale and PDF, are taken into account), however, we prefer to present
the predictions separately.

In addition, we have also presented cross sections for the production
of a fourth generation $b^\prime$, both in association with a top
quark and with its partner $t^\prime$. These cross sections set useful
benchmarks for future searches, particularly at the LHC where very
heavy quarks with sizeable mixing with third generation quarks or very
large mass splittings would be preferentially produced from $t$-channel production
rather than in pairs via the strong interaction.

Although the cross sections presented here embody the current
state-of-the-art, a number of avenues for future refinement are
evident. First, given the importance of threshold resummation in both
the $s$-channel and pair production modes, the $t$-channel predictions
here could be further improved by including such effects. This would
be particularly important at the Tevatron and for high mass $t^\prime$
production at the LHC. Second, in the near future a calculation of the
$2 \to 2$ process at NNLO should be feasible. With such a calculation
one would be able to better assess the importance of higher order
effects in the strong coupling as well as the resummation of collinear
logarithms in the bottom quark PDF.
This could help to confirm whether these logarithms are large enough that
their resummation leads to an important effect, which does not appear to be
the case given the reasonable agreement between the $2 \to 2$ and $2 \to 3$
NLO calculations presented here.

\section*{Acknowledgements}
We are grateful to Fred Olness for providing us with the means to 
generate PDFs for different bottom quark masses and to Pavel Demin and
the IT staff of CP3 for the computing support. 

\bibliography{physics}


\appendix
\section{Cross sections and uncertainties for single top and $t^\prime$ production}\label{app:A}

\begin{small}
\renewcommand{\arraystretch}{1.1}
\TABULAR[h]{cr@{ }lr@{ }l}
{
\toprule[0.08em]
top mass (GeV)&\multicolumn{2}{c}{{$2 \to 2$} (fb)} &\multicolumn{2}{c}{{$2 \to 3$} (fb)}\\
\midrule[0.05em]
  164&$(  1059 )$&$  1127  ^{+    27 }_{-     2}~^{+   106 }_{-    89}$&$(   790 )$&$  1080  ^{+    88 }_{-   121}~^{+    96 }_{-    80 }$\\
  168&$(   985 )$&$  1052  ^{+    23 }_{-     3}~^{+   102 }_{-    86}$&$(   729 )$&$  1006  ^{+    83 }_{-   115}~^{+    93 }_{-    77 }$\\
  172&$(   917 )$&$   981  ^{+    23 }_{-     3}~^{+    98 }_{-    82}$&$(   672 )$&$   935  ^{+    82 }_{-   107}~^{+    88 }_{-    74 }$\\
  176&$(   854 )$&$   916  ^{+    21 }_{-     2}~^{+    95 }_{-    79}$&$(   621 )$&$   872  ^{+    76 }_{-   101}~^{+    85 }_{-    71 }$\\
  180&$(   796 )$&$   856  ^{+    20 }_{-     3}~^{+    91 }_{-    76}$&$(   575 )$&$   812  ^{+    74 }_{-    95}~^{+    82 }_{-    68 }$\\
  200&$(   561 )$&$   613  ^{+    13 }_{-     2}~^{+    75 }_{-    62}$&$(   391 )$&$   575  ^{+    58 }_{-    71}~^{+    66 }_{-    54 }$\\
  400&$( 23.7 )$&$ 33.5  ^{+  0.4 }_{-  0.3}~^{+ 10.4 }_{-  8.0}$&$( 13.1 )$&$ 27.9  ^{+  4.9 }_{-  4.7}~^{+  7.9 }_{-  6.0 }$\\
  600&$( 1.24 )$&$ 2.57  ^{+ 0.05 }_{- 0.07}~^{+ 1.38 }_{- 0.99}$&$( 0.59 )$&$ 1.88  ^{+ 0.45 }_{- 0.39}~^{+ 0.90 }_{- 0.64 }$\\
  800&$(0.064 )$&$0.202  ^{+0.009 }_{-0.010}~^{+0.155 }_{-0.107}$&$(0.026 )$&$0.126  ^{+0.039 }_{-0.031}~^{+0.088 }_{-0.060 }$\\
 1000&$(0.003 )$&$0.013  ^{+0.001 }_{-0.001}~^{+0.013 }_{-0.008}$&$(0.001 )$&$0.007  ^{+0.002 }_{-0.002}~^{+0.006 }_{-0.004 }$\\
 
\bottomrule[0.08em]
 }
{ 
\label{tab:1}
NLO cross sections (fb) at the Tevatron Run II for top quark
production in the $t$ channel, as a function of the top mass obtained
with the CTEQ6.6 PDF set and $V_{t^{(\prime)}b}=1$. In the second
(third) column the $2\to2$ ($2\to3$) results are shown, where the
first uncertainty comes from renormalisation and factorisation scales
variation and the second from PDF errors. Numbers in parenthesis refer
to the corresponding LO results. Cross sections for anti-top are the
same and are not displayed.  These results are plotted in
Fig.~\ref{fig:1} where the scale and PDF uncertainties are combined
linearly.  }
\end{small}


\begin{small}
\renewcommand{\arraystretch}{1.1}
\TABULAR[h]{cr@{ }lr@{ }l}
{
\toprule[0.08em]
top mass (GeV)&\multicolumn{2}{c}{{$2 \to 2$} (fb)} &\multicolumn{2}{c}{{$2 \to 3$} (fb)}\\
\midrule[0.05em]
  164&$(  1211 )$&$  1145  ^{+    28 }_{-     2}~^{+    66 }_{-    57}$&$(   893 )$&$  1089  ^{+    94 }_{-   127}~^{+    58 }_{-    47 }$\\
  168&$(  1131 )$&$  1067  ^{+    27 }_{-     2}~^{+    63 }_{-    54}$&$(   827 )$&$  1013  ^{+    88 }_{-   122}~^{+    56 }_{-    45 }$\\
  172&$(  1056 )$&$   994  ^{+    24 }_{-     2}~^{+    61 }_{-    52}$&$(   764 )$&$   942  ^{+    86 }_{-   113}~^{+    53 }_{-    43 }$\\
  176&$(   987 )$&$   928  ^{+    23 }_{-     3}~^{+    58 }_{-    50}$&$(   707 )$&$   876  ^{+    80 }_{-   107}~^{+    51 }_{-    41 }$\\
  180&$(   922 )$&$   866  ^{+    21 }_{-     3}~^{+    56 }_{-    47}$&$(   655 )$&$   816  ^{+    75 }_{-   100}~^{+    49 }_{-    39 }$\\
  200&$(   661 )$&$   618  ^{+    14 }_{-     3}~^{+    45 }_{-    38}$&$(   450 )$&$   574  ^{+    61 }_{-    74}~^{+    39 }_{-    31 }$\\
  400&$( 33.0 )$&$ 30.0  ^{+  0.4 }_{-  0.3}~^{+  5.1 }_{-  4.0}$&$( 16.6 )$&$ 24.9  ^{+  4.7 }_{-  4.4}~^{+  3.7 }_{-  2.8 }$\\
  600&$(2.032 )$&$1.811  ^{+0.040 }_{-0.060}~^{+0.512 }_{-0.389}$&$(0.813 )$&$1.352  ^{+0.342 }_{-0.290}~^{+0.323 }_{-0.235 }$\\
  800&$(0.116 )$&$0.100  ^{+0.005 }_{-0.005}~^{+0.042 }_{-0.030}$&$(0.038 )$&$0.067  ^{+0.022 }_{-0.016}~^{+0.023 }_{-0.016 }$\\
 1000&$(0.005 )$&$0.004  ^{+0.001 }_{-0.000}~^{+0.002 }_{-0.002}$&$(0.001 )$&$0.003  ^{+0.000 }_{-0.001}~^{+0.001 }_{-0.001 }$\\
 
\bottomrule[0.08em]
 }
{ 
\label{tab:2}
Same as Table~\protect\ref{tab:1} but for the MSTW2008 PDF set.
}
\end{small}

\begin{small}
\renewcommand{\arraystretch}{1.1}
\TABULAR[h]{crr@{ }lr@{ }l}
{
\toprule[0.08em]
\multicolumn{2}{c}{top mass (GeV)}&\multicolumn{2}{c}{{$2 \to 2$} (pb)} &\multicolumn{2}{c}{{$2 \to 3$} (pb)}\\
\midrule[0.05em]
  \multirow{2}{2cm}{  164 } & $t$ &$( 80.9 )$&$ 89.5  ^{+  1.5 }_{-  1.1}~^{+  1.7 }_{-  1.8}$&$( 80.9 )$&$ 86.0  ^{+  2.6 }_{-  3.8}~^{+  1.6 }_{-  1.7 }$\\
 \vspace{3pt}
      & $\bar t$ &$( 44.1 )$&$ 50.2  ^{+  0.8 }_{-  0.5}~^{+  1.1 }_{-  1.2}$&$( 43.3 )$&$ 47.7  ^{+  1.4 }_{-  2.0}~^{+  1.9 }_{-  2.2 }$\\
  \multirow{2}{2cm}{  168 } & $t$ &$( 78.2 )$&$ 86.4  ^{+  1.6 }_{-  1.0}~^{+  1.6 }_{-  1.8}$&$( 77.7 )$&$ 83.0  ^{+  2.4 }_{-  3.9}~^{+  1.5 }_{-  1.6 }$\\
 \vspace{3pt}
      & $\bar t$ &$( 42.6 )$&$ 48.4  ^{+  0.7 }_{-  0.6}~^{+  1.0 }_{-  1.2}$&$( 41.6 )$&$ 46.0  ^{+  1.2 }_{-  2.1}~^{+  1.0 }_{-  1.1 }$\\
  \multirow{2}{2cm}{  172 } & $t$ &$( 75.6 )$&$ 83.5  ^{+  1.4 }_{-  1.1}~^{+  1.5 }_{-  1.7}$&$( 74.8 )$&$ 79.8  ^{+  2.9 }_{-  3.4}~^{+  1.4 }_{-  1.6 }$\\
 \vspace{3pt}
      & $\bar t$ &$( 41.1 )$&$ 46.6  ^{+  0.8 }_{-  0.5}~^{+  1.0 }_{-  1.1}$&$( 39.8 )$&$ 44.2  ^{+  1.2 }_{-  2.0}~^{+  1.0 }_{-  1.1 }$\\
  \multirow{2}{2cm}{  176 } & $t$ &$( 73.1 )$&$ 80.6  ^{+  1.4 }_{-  0.9}~^{+  1.5 }_{-  1.6}$&$( 71.9 )$&$ 77.2  ^{+  2.6 }_{-  3.8}~^{+  1.4 }_{-  1.5 }$\\
 \vspace{3pt}
      & $\bar t$ &$( 39.7 )$&$ 44.9  ^{+  0.7 }_{-  0.5}~^{+  1.0 }_{-  1.1}$&$( 38.3 )$&$ 42.4  ^{+  1.3 }_{-  1.9}~^{+  0.9 }_{-  1.1 }$\\
  \multirow{2}{2cm}{  180 } & $t$ &$( 70.8 )$&$ 77.9  ^{+  1.3 }_{-  0.8}~^{+  1.4 }_{-  1.6}$&$( 68.8 )$&$ 74.6  ^{+  2.3 }_{-  3.7}~^{+  1.4 }_{-  1.4 }$\\
 \vspace{3pt}
      & $\bar t$ &$( 38.3 )$&$ 43.3  ^{+  0.8 }_{-  0.4}~^{+  0.9 }_{-  1.1}$&$( 36.6 )$&$ 40.9  ^{+  1.5 }_{-  1.8}~^{+  0.9 }_{-  1.0 }$\\
  \multirow{2}{2cm}{  200 } & $t$ &$( 60.1 )$&$ 66.1  ^{+  1.1 }_{-  0.7}~^{+  1.2 }_{-  1.3}$&$( 57.1 )$&$ 62.6  ^{+  2.6 }_{-  3.1}~^{+  1.2 }_{-  1.2 }$\\
 \vspace{3pt}
      & $\bar t$ &$( 32.3 )$&$ 36.4  ^{+  0.6 }_{-  0.3}~^{+  0.8 }_{-  0.9}$&$( 30.0 )$&$ 33.8  ^{+  1.5 }_{-  1.4}~^{+  0.8 }_{-  0.9 }$\\
  \multirow{2}{2cm}{  400 } & $t$ &$(15.86 )$&$17.65  ^{+ 0.24 }_{- 0.08}~^{+ 0.53 }_{- 0.52}$&$(12.65 )$&$16.13  ^{+ 0.93 }_{- 1.21}~^{+ 0.49 }_{- 0.47 }$\\
 \vspace{3pt}
      & $\bar t$ &$( 7.87 )$&$ 8.99  ^{+ 0.13 }_{- 0.04}~^{+ 0.41 }_{- 0.43}$&$( 6.13 )$&$ 8.01  ^{+ 0.50 }_{- 0.56}~^{+ 0.39 }_{- 0.40 }$\\
  \multirow{2}{2cm}{  600 } & $t$ &$( 5.65 )$&$ 6.46  ^{+ 0.07 }_{- 0.03}~^{+ 0.35 }_{- 0.31}$&$( 4.09 )$&$ 5.68  ^{+ 0.46 }_{- 0.50}~^{+ 0.31 }_{- 0.27 }$\\
 \vspace{3pt}
      & $\bar t$ &$( 2.62 )$&$ 3.10  ^{+ 0.04 }_{- 0.01}~^{+ 0.23 }_{- 0.23}$&$( 1.84 )$&$ 2.66  ^{+ 0.20 }_{- 0.24}~^{+ 0.21 }_{- 0.20 }$\\
  \multirow{2}{2cm}{  800 } & $t$ &$( 2.35 )$&$ 2.77  ^{+ 0.03 }_{- 0.02}~^{+ 0.23 }_{- 0.19}$&$( 1.58 )$&$ 2.39  ^{+ 0.22 }_{- 0.25}~^{+ 0.19 }_{- 0.16 }$\\
 \vspace{3pt}
      & $\bar t$ &$(1.025 )$&$1.263  ^{+0.012 }_{-0.009}~^{+0.138 }_{-0.127}$&$(0.672 )$&$1.053  ^{+0.093 }_{-0.104}~^{+0.116 }_{-0.106 }$\\
  \multirow{2}{2cm}{ 1000 } & $t$ &$(1.071 )$&$1.311  ^{+0.019 }_{-0.012}~^{+0.146 }_{-0.121}$&$(0.680 )$&$1.093  ^{+0.121 }_{-0.121}~^{+0.118 }_{-0.096 }$\\
 \vspace{3pt}
      & $\bar t$ &$(0.442 )$&$0.570  ^{+0.006 }_{-0.005}~^{+0.082 }_{-0.073}$&$(0.274 )$&$0.462  ^{+0.043 }_{-0.051}~^{+0.066 }_{-0.059 }$\\
  \multirow{2}{2cm}{ 1200 } & $t$ &$(0.518 )$&$0.660  ^{+0.010 }_{-0.006}~^{+0.095 }_{-0.077}$&$(0.315 )$&$0.537  ^{+0.068 }_{-0.062}~^{+0.073 }_{-0.059 }$\\
 \vspace{3pt}
      & $\bar t$ &$(0.203 )$&$0.275  ^{+0.004 }_{-0.003}~^{+0.050 }_{-0.043}$&$(0.120 )$&$0.217  ^{+0.026 }_{-0.026}~^{+0.038 }_{-0.033 }$\\
  \multirow{2}{2cm}{ 1400 } & $t$ &$(0.262 )$&$0.348  ^{+0.006 }_{-0.005}~^{+0.062 }_{-0.049}$&$(0.153 )$&$0.276  ^{+0.037 }_{-0.033}~^{+0.045 }_{-0.036 }$\\
 \vspace{3pt}
      & $\bar t$ &$(0.097 )$&$0.140  ^{+0.002 }_{-0.002}~^{+0.031 }_{-0.026}$&$(0.056 )$&$0.108  ^{+0.013 }_{-0.013}~^{+0.023 }_{-0.019 }$\\
  \multirow{2}{2cm}{ 1600 } & $t$ &$(0.136 )$&$0.190  ^{+0.004 }_{-0.004}~^{+0.041 }_{-0.032}$&$(0.077 )$&$0.148  ^{+0.021 }_{-0.020}~^{+0.029 }_{-0.022 }$\\
 \vspace{3pt}
      & $\bar t$ &$(0.048 )$&$0.074  ^{+0.001 }_{-0.001}~^{+0.019 }_{-0.016}$&$(0.027 )$&$0.055  ^{+0.008 }_{-0.007}~^{+0.014 }_{-0.011 }$\\
  \multirow{2}{2cm}{ 1800 } & $t$ &$(0.072 )$&$0.107  ^{+0.003 }_{-0.002}~^{+0.027 }_{-0.021}$&$(0.040 )$&$0.081  ^{+0.012 }_{-0.011}~^{+0.018 }_{-0.014 }$\\
 \vspace{3pt}
      & $\bar t$ &$(0.025 )$&$0.040  ^{+0.001 }_{-0.001}~^{+0.012 }_{-0.010}$&$(0.013 )$&$0.029  ^{+0.004 }_{-0.004}~^{+0.008 }_{-0.007 }$\\
  \multirow{2}{2cm}{ 2000 } & $t$ &$(0.039 )$&$0.061  ^{+0.002 }_{-0.002}~^{+0.018 }_{-0.014}$&$(0.021 )$&$0.045  ^{+0.007 }_{-0.006}~^{+0.012 }_{-0.009 }$\\
 \vspace{3pt}
      & $\bar t$ &$(0.013 )$&$0.022  ^{+0.001 }_{-0.001}~^{+0.008 }_{-0.006}$&$(0.007 )$&$0.016  ^{+0.003 }_{-0.002}~^{+0.005 }_{-0.004 }$\\
 
\bottomrule[0.08em]
 }
{ 
\label{tab:3}
NLO cross sections (pb) at the LHC 10 TeV for top and anti-top quarks
production, as a function of the top mass obtained with the CTEQ6.6
PDF set and $V_{t^{(\prime)}b}=1$. In the second (third) column the
$2\to2$ ($2\to3$) results are shown, where the first uncertainty comes
from renormalisation and factorisation scales variation and the second
from PDF errors. Numbers in parenthesis refer to the corresponding LO
results.  These results are plotted in Fig.~\ref{fig:3} where the
scale and PDF uncertainties are combined linearly.  }
\end{small}

\begin{small}
\renewcommand{\arraystretch}{1.1}
\TABULAR[h]{crr@{ }lr@{ }l}
{
\toprule[0.08em]
\multicolumn{2}{c}{top mass (GeV)}&\multicolumn{2}{c}{{$2 \to 2$} (pb)} &\multicolumn{2}{c}{{$2 \to 3$} (pb)}\\
\midrule[0.05em]
  \multirow{2}{2cm}{  164 } & $t$ &$( 82.0 )$&$ 90.4  ^{+  1.6 }_{-  1.1}~^{+  1.2 }_{-  1.1}$&$( 82.3 )$&$ 86.7  ^{+  2.6 }_{-  4.0}~^{+  1.2 }_{-  1.0 }$\\
 \vspace{3pt}
      & $\bar t$ &$( 46.4 )$&$ 52.1  ^{+  0.9 }_{-  0.6}~^{+  0.8 }_{-  1.1}$&$( 45.8 )$&$ 48.7  ^{+  1.8 }_{-  1.8}~^{+  0.7 }_{-  1.0 }$\\
  \multirow{2}{2cm}{  168 } & $t$ &$( 79.2 )$&$ 87.3  ^{+  1.6 }_{-  1.1}~^{+  1.1 }_{-  1.1}$&$( 79.1 )$&$ 83.5  ^{+  2.7 }_{-  4.0}~^{+  1.2 }_{-  1.0 }$\\
 \vspace{3pt}
      & $\bar t$ &$( 44.7 )$&$ 50.2  ^{+  0.9 }_{-  0.6}~^{+  0.8 }_{-  1.0}$&$( 43.8 )$&$ 47.3  ^{+  1.1 }_{-  2.2}~^{+  0.8 }_{-  1.0 }$\\
  \multirow{2}{2cm}{  172 } & $t$ &$( 76.6 )$&$ 84.4  ^{+  1.4 }_{-  1.0}~^{+  1.1 }_{-  1.0}$&$( 76.0 )$&$ 80.3  ^{+  3.2 }_{-  3.7}~^{+  1.1 }_{-  0.9 }$\\
 \vspace{3pt}
      & $\bar t$ &$( 43.2 )$&$ 48.3  ^{+  0.8 }_{-  0.5}~^{+  0.7 }_{-  1.0}$&$( 42.0 )$&$ 45.4  ^{+  1.7 }_{-  2.1}~^{+  0.7 }_{-  0.9 }$\\
  \multirow{2}{2cm}{  176 } & $t$ &$( 74.1 )$&$ 81.6  ^{+  1.4 }_{-  0.9}~^{+  1.1 }_{-  1.0}$&$( 73.1 )$&$ 77.6  ^{+  2.3 }_{-  3.4}~^{+  1.1 }_{-  0.9 }$\\
 \vspace{3pt}
      & $\bar t$ &$( 41.7 )$&$ 46.6  ^{+  0.8 }_{-  0.5}~^{+  0.7 }_{-  1.0}$&$( 40.3 )$&$ 43.7  ^{+  1.2 }_{-  1.9}~^{+  0.7 }_{-  0.9 }$\\
  \multirow{2}{2cm}{  180 } & $t$ &$( 71.7 )$&$ 78.8  ^{+  1.4 }_{-  0.8}~^{+  1.0 }_{-  0.9}$&$( 70.3 )$&$ 74.6  ^{+  3.0 }_{-  3.2}~^{+  1.0 }_{-  0.9 }$\\
 \vspace{3pt}
      & $\bar t$ &$( 40.3 )$&$ 45.0  ^{+  0.7 }_{-  0.5}~^{+  0.7 }_{-  0.9}$&$( 38.6 )$&$ 42.3  ^{+  1.0 }_{-  2.2}~^{+  0.7 }_{-  0.9 }$\\
  \multirow{2}{2cm}{  200 } & $t$ &$( 61.1 )$&$ 67.0  ^{+  1.1 }_{-  0.6}~^{+  0.9 }_{-  0.8}$&$( 58.2 )$&$ 63.3  ^{+  2.6 }_{-  3.2}~^{+  0.9 }_{-  0.7 }$\\
 \vspace{3pt}
      & $\bar t$ &$( 34.1 )$&$ 37.8  ^{+  0.6 }_{-  0.4}~^{+  0.6 }_{-  0.8}$&$( 31.7 )$&$ 35.3  ^{+  0.9 }_{-  1.7}~^{+  0.6 }_{-  0.8 }$\\
  \multirow{2}{2cm}{  400 } & $t$ &$(16.43 )$&$18.02  ^{+ 0.22 }_{- 0.10}~^{+ 0.38 }_{- 0.30}$&$(13.02 )$&$16.37  ^{+ 0.98 }_{- 1.31}~^{+ 0.35 }_{- 0.26 }$\\
 \vspace{3pt}
      & $\bar t$ &$( 8.51 )$&$ 9.33  ^{+ 0.13 }_{- 0.06}~^{+ 0.29 }_{- 0.32}$&$( 6.55 )$&$ 8.33  ^{+ 0.45 }_{- 0.65}~^{+ 0.26 }_{- 0.28 }$\\
  \multirow{2}{2cm}{  600 } & $t$ &$( 6.01 )$&$ 6.60  ^{+ 0.08 }_{- 0.03}~^{+ 0.23 }_{- 0.19}$&$( 4.27 )$&$ 5.78  ^{+ 0.48 }_{- 0.53}~^{+ 0.20 }_{- 0.15 }$\\
 \vspace{3pt}
      & $\bar t$ &$( 2.90 )$&$ 3.19  ^{+ 0.04 }_{- 0.02}~^{+ 0.15 }_{- 0.16}$&$( 1.99 )$&$ 2.74  ^{+ 0.21 }_{- 0.26}~^{+ 0.16 }_{- 0.16 }$\\
  \multirow{2}{2cm}{  800 } & $t$ &$( 2.57 )$&$ 2.82  ^{+ 0.03 }_{- 0.02}~^{+ 0.14 }_{- 0.11}$&$( 1.68 )$&$ 2.40  ^{+ 0.24 }_{- 0.25}~^{+ 0.12 }_{- 0.09 }$\\
 \vspace{3pt}
      & $\bar t$ &$(1.167 )$&$1.281  ^{+0.012 }_{-0.010}~^{+0.084 }_{-0.082}$&$(0.734 )$&$1.064  ^{+0.092 }_{-0.114}~^{+0.068 }_{-0.064 }$\\
  \multirow{2}{2cm}{ 1000 } & $t$ &$(1.207 )$&$1.324  ^{+0.015 }_{-0.014}~^{+0.088 }_{-0.070}$&$(0.733 )$&$1.091  ^{+0.127 }_{-0.124}~^{+0.069 }_{-0.050 }$\\
 \vspace{3pt}
      & $\bar t$ &$(0.516 )$&$0.568  ^{+0.007 }_{-0.005}~^{+0.048 }_{-0.045}$&$(0.303 )$&$0.453  ^{+0.052 }_{-0.051}~^{+0.037 }_{-0.033 }$\\
  \multirow{2}{2cm}{ 1200 } & $t$ &$(0.602 )$&$0.658  ^{+0.011 }_{-0.008}~^{+0.055 }_{-0.043}$&$(0.344 )$&$0.529  ^{+0.070 }_{-0.066}~^{+0.041 }_{-0.030 }$\\
 \vspace{3pt}
      & $\bar t$ &$(0.244 )$&$0.268  ^{+0.004 }_{-0.003}~^{+0.028 }_{-0.025}$&$(0.134 )$&$0.209  ^{+0.025 }_{-0.027}~^{+0.021 }_{-0.018 }$\\
  \multirow{2}{2cm}{ 1400 } & $t$ &$(0.314 )$&$0.342  ^{+0.006 }_{-0.006}~^{+0.034 }_{-0.027}$&$(0.170 )$&$0.268  ^{+0.039 }_{-0.034}~^{+0.024 }_{-0.018 }$\\
 \vspace{3pt}
      & $\bar t$ &$(0.120 )$&$0.132  ^{+0.003 }_{-0.002}~^{+0.017 }_{-0.015}$&$(0.063 )$&$0.100  ^{+0.014 }_{-0.013}~^{+0.012 }_{-0.010 }$\\
  \multirow{2}{2cm}{ 1600 } & $t$ &$(0.168 )$&$0.183  ^{+0.004 }_{-0.003}~^{+0.022 }_{-0.017}$&$(0.086 )$&$0.141  ^{+0.021 }_{-0.020}~^{+0.015 }_{-0.011 }$\\
 \vspace{3pt}
      & $\bar t$ &$(0.061 )$&$0.068  ^{+0.001 }_{-0.002}~^{+0.010 }_{-0.009}$&$(0.030 )$&$0.050  ^{+0.008 }_{-0.007}~^{+0.007 }_{-0.006 }$\\
  \multirow{2}{2cm}{ 1800 } & $t$ &$(0.092 )$&$0.100  ^{+0.003 }_{-0.002}~^{+0.014 }_{-0.011}$&$(0.045 )$&$0.075  ^{+0.013 }_{-0.011}~^{+0.009 }_{-0.006 }$\\
 \vspace{3pt}
      & $\bar t$ &$(0.032 )$&$0.035  ^{+0.001 }_{-0.001}~^{+0.006 }_{-0.005}$&$(0.015 )$&$0.025  ^{+0.004 }_{-0.004}~^{+0.004 }_{-0.003 }$\\
  \multirow{2}{2cm}{ 2000 } & $t$ &$(0.051 )$&$0.056  ^{+0.002 }_{-0.002}~^{+0.009 }_{-0.007}$&$(0.024 )$&$0.041  ^{+0.007 }_{-0.006}~^{+0.005 }_{-0.004 }$\\
 \vspace{3pt}
      & $\bar t$ &$(0.017 )$&$0.019  ^{+0.001 }_{-0.001}~^{+0.004 }_{-0.003}$&$(0.008 )$&$0.013  ^{+0.002 }_{-0.002}~^{+0.002 }_{-0.002 }$\\
 
\bottomrule[0.08em]
 }
{ 
\label{tab:4}
Same as Table~\protect\ref{tab:3} but for the MSTW2008 PDF set.
}
\end{small}

 \begin{small}
 \renewcommand{\arraystretch}{1.1}
 \TABULAR[h]{crr@{ }lr@{ }l}
 {
 \toprule[0.08em]
 \multicolumn{2}{c}{top mass (GeV)}&\multicolumn{2}{c}{{$2 \to 2$} (pb)} &\multicolumn{2}{c}{{$2 \to 3$} (pb)}\\
 \midrule[0.05em]
   \multirow{2}{2cm}{  164 } & $t$ &$(146.1 )$&$163.2  ^{+  2.8 }_{-  2.7}~^{+  3.2 }_{-  3.7}$&$(152.3 )$&$156.5  ^{+  5.2 }_{-  5.9}~^{+  3.0 }_{-  3.4 }$\\
 \vspace{3pt}
      & $\bar t$ &$( 85.1 )$&$ 97.3  ^{+  1.7 }_{-  1.5}~^{+  1.9 }_{-  2.2}$&$( 86.7 )$&$ 92.8  ^{+  2.6 }_{-  3.6}~^{+  1.8 }_{-  2.0 }$\\
  \multirow{2}{2cm}{  168 } & $t$ &$(141.9 )$&$158.2  ^{+  2.7 }_{-  2.6}~^{+  3.1 }_{-  3.6}$&$(146.7 )$&$151.9  ^{+  5.0 }_{-  5.7}~^{+  2.9 }_{-  3.3 }$\\
 \vspace{3pt}
      & $\bar t$ &$( 82.4 )$&$ 94.1  ^{+  1.6 }_{-  1.5}~^{+  1.9 }_{-  2.2}$&$( 83.4 )$&$ 89.8  ^{+  2.8 }_{-  3.6}~^{+  1.8 }_{-  2.0 }$\\
  \multirow{2}{2cm}{  172 } & $t$ &$(137.6 )$&$152.9  ^{+  3.0 }_{-  2.3}~^{+  3.0 }_{-  3.4}$&$(141.8 )$&$147.0  ^{+  5.0 }_{-  5.7}~^{+  2.7 }_{-  3.1 }$\\
 \vspace{3pt}
      & $\bar t$ &$( 79.8 )$&$ 91.1  ^{+  1.5 }_{-  1.5}~^{+  1.8 }_{-  2.1}$&$( 80.5 )$&$ 86.8  ^{+  2.4 }_{-  3.7}~^{+  1.8 }_{-  2.0 }$\\
  \multirow{2}{2cm}{  176 } & $t$ &$(133.6 )$&$148.5  ^{+  2.6 }_{-  2.3}~^{+  2.9 }_{-  3.3}$&$(136.8 )$&$142.6  ^{+  4.9 }_{-  5.6}~^{+  2.7 }_{-  3.0 }$\\
 \vspace{3pt}
      & $\bar t$ &$( 77.5 )$&$ 88.1  ^{+  1.5 }_{-  1.3}~^{+  1.7 }_{-  2.0}$&$( 77.8 )$&$ 83.6  ^{+  2.7 }_{-  3.0}~^{+  1.6 }_{-  1.8 }$\\
  \multirow{2}{2cm}{  180 } & $t$ &$(129.6 )$&$144.0  ^{+  2.5 }_{-  2.2}~^{+  2.8 }_{-  3.2}$&$(131.6 )$&$138.1  ^{+  4.9 }_{-  5.3}~^{+  2.6 }_{-  2.9 }$\\
 \vspace{3pt}
      & $\bar t$ &$( 75.0 )$&$ 85.3  ^{+  1.5 }_{-  1.2}~^{+  1.7 }_{-  2.0}$&$( 74.6 )$&$ 81.1  ^{+  2.1 }_{-  3.6}~^{+  1.6 }_{-  1.8 }$\\
  \multirow{2}{2cm}{  200 } & $t$ &$(112.2 )$&$124.2  ^{+  2.2 }_{-  1.7}~^{+  2.3 }_{-  2.6}$&$(111.1 )$&$117.0  ^{+  4.4 }_{-  3.8}~^{+  2.1 }_{-  2.4 }$\\
 \vspace{3pt}
      & $\bar t$ &$( 64.5 )$&$ 72.9  ^{+  1.3 }_{-  1.0}~^{+  1.5 }_{-  1.7}$&$( 62.4 )$&$ 68.9  ^{+  2.1 }_{-  2.8}~^{+  1.4 }_{-  1.6 }$\\
  \multirow{2}{2cm}{  400 } & $t$ &$( 34.5 )$&$ 38.3  ^{+  0.5 }_{-  0.3}~^{+  0.8 }_{-  0.8}$&$( 28.9 )$&$ 35.2  ^{+  1.9 }_{-  2.0}~^{+  0.8 }_{-  0.8 }$\\
 \vspace{3pt}
      & $\bar t$ &$( 18.4 )$&$ 20.9  ^{+  0.3 }_{-  0.1}~^{+  0.6 }_{-  0.7}$&$( 15.1 )$&$ 19.1  ^{+  0.7 }_{-  1.3}~^{+  0.6 }_{-  0.6 }$\\
  \multirow{2}{2cm}{  600 } & $t$ &$(14.01 )$&$15.92  ^{+ 0.17 }_{- 0.09}~^{+ 0.53 }_{- 0.51}$&$(10.71 )$&$14.25  ^{+ 0.94 }_{- 1.18}~^{+ 0.48 }_{- 0.46 }$\\
 \vspace{3pt}
      & $\bar t$ &$( 7.06 )$&$ 8.20  ^{+ 0.10 }_{- 0.04}~^{+ 0.41 }_{- 0.42}$&$( 5.25 )$&$ 7.18  ^{+ 0.44 }_{- 0.56}~^{+ 0.37 }_{- 0.38 }$\\
  \multirow{2}{2cm}{  800 } & $t$ &$( 6.60 )$&$ 7.66  ^{+ 0.08 }_{- 0.05}~^{+ 0.39 }_{- 0.35}$&$( 4.71 )$&$ 6.64  ^{+ 0.60 }_{- 0.55}~^{+ 0.34 }_{- 0.30 }$\\
 \vspace{3pt}
      & $\bar t$ &$( 3.15 )$&$ 3.77  ^{+ 0.04 }_{- 0.02}~^{+ 0.27 }_{- 0.26}$&$( 2.19 )$&$ 3.20  ^{+ 0.25 }_{- 0.26}~^{+ 0.24 }_{- 0.23 }$\\
  \multirow{2}{2cm}{ 1000 } & $t$ &$( 3.40 )$&$ 4.05  ^{+ 0.04 }_{- 0.04}~^{+ 0.28 }_{- 0.24}$&$( 2.30 )$&$ 3.43  ^{+ 0.33 }_{- 0.31}~^{+ 0.24 }_{- 0.20 }$\\
 \vspace{3pt}
      & $\bar t$ &$(1.546 )$&$1.907  ^{+0.019 }_{-0.014}~^{+0.179 }_{-0.169}$&$(1.020 )$&$1.585  ^{+0.129 }_{-0.143}~^{+0.152 }_{-0.142 }$\\
  \multirow{2}{2cm}{ 1200 } & $t$ &$( 1.86 )$&$ 2.27  ^{+ 0.03 }_{- 0.02}~^{+ 0.21 }_{- 0.17}$&$( 1.21 )$&$ 1.90  ^{+ 0.19 }_{- 0.19}~^{+ 0.17 }_{- 0.14 }$\\
 \vspace{3pt}
      & $\bar t$ &$(0.810 )$&$1.033  ^{+0.009 }_{-0.009}~^{+0.123 }_{-0.112}$&$(0.513 )$&$0.842  ^{+0.070 }_{-0.086}~^{+0.100 }_{-0.091 }$\\
  \multirow{2}{2cm}{ 1400 } & $t$ &$(1.064 )$&$1.333  ^{+0.016 }_{-0.016}~^{+0.149 }_{-0.123}$&$(0.663 )$&$1.090  ^{+0.124 }_{-0.117}~^{+0.118 }_{-0.096 }$\\
 \vspace{3pt}
      & $\bar t$ &$(0.444 )$&$0.585  ^{+0.006 }_{-0.006}~^{+0.085 }_{-0.075}$&$(0.270 )$&$0.470  ^{+0.043 }_{-0.053}~^{+0.067 }_{-0.060 }$\\
  \multirow{2}{2cm}{ 1600 } & $t$ &$(0.626 )$&$0.807  ^{+0.012 }_{-0.010}~^{+0.109 }_{-0.089}$&$(0.378 )$&$0.651  ^{+0.074 }_{-0.077}~^{+0.083 }_{-0.067 }$\\
 \vspace{3pt}
      & $\bar t$ &$(0.252 )$&$0.343  ^{+0.006 }_{-0.003}~^{+0.059 }_{-0.051}$&$(0.149 )$&$0.268  ^{+0.034 }_{-0.031}~^{+0.044 }_{-0.039 }$\\
  \multirow{2}{2cm}{ 1800 } & $t$ &$(0.378 )$&$0.503  ^{+0.008 }_{-0.008}~^{+0.080 }_{-0.064}$&$(0.221 )$&$0.396  ^{+0.051 }_{-0.046}~^{+0.059 }_{-0.047 }$\\
 \vspace{3pt}
      & $\bar t$ &$(0.146 )$&$0.208  ^{+0.004 }_{-0.003}~^{+0.041 }_{-0.035}$&$(0.084 )$&$0.159  ^{+0.020 }_{-0.019}~^{+0.030 }_{-0.026 }$\\
  \multirow{2}{2cm}{ 2000 } & $t$ &$(0.233 )$&$0.319  ^{+0.007 }_{-0.005}~^{+0.059 }_{-0.047}$&$(0.133 )$&$0.251  ^{+0.030 }_{-0.033}~^{+0.042 }_{-0.033 }$\\
 \vspace{3pt}
      & $\bar t$ &$(0.087 )$&$0.128  ^{+0.002 }_{-0.002}~^{+0.029 }_{-0.024}$&$(0.049 )$&$0.096  ^{+0.014 }_{-0.012}~^{+0.021 }_{-0.018 }$\\
 
 \bottomrule[0.08em]
  }
 { 
 \label{tab:5}
 NLO cross sections (pb) at the LHC 14 TeV for top and anti-top quarks
 production, as a function of the top mass obtained with the CTEQ6.6
 PDF set and $V_{t^{(\prime)}b}=1$. In the second (third) column the
 $2\to2$ ($2\to3$) results are shown, where the first uncertainty
 comes from renormalisation and factorisation scales variation and the
 second from PDF errors. Numbers in parenthesis refer to the
 corresponding LO results.  These results are plotted in
 Fig.~\ref{fig:5} where the scale and PDF uncertainties are combined
 linearly.  }
 \end{small}

 \begin{small}
 \renewcommand{\arraystretch}{1.1}
 \TABULAR[h]{crr@{ }lr@{ }l}
 {
 \toprule[0.08em]
 \multicolumn{2}{c}{top mass (GeV)}&\multicolumn{2}{c}{{$2 \to 2$} (pb)} &\multicolumn{2}{c}{{$2 \to 3$} (pb)}\\
 \midrule[0.05em]
   \multirow{2}{2cm}{  164 } & $t$ &$(148.1 )$&$164.1  ^{+  3.1 }_{-  2.8}~^{+  2.4 }_{-  2.4}$&$(155.1 )$&$157.8  ^{+  4.4 }_{-  6.9}~^{+  2.4 }_{-  2.3 }$\\
 \vspace{3pt}
      & $\bar t$ &$( 88.8 )$&$100.6  ^{+  1.7 }_{-  1.8}~^{+  1.3 }_{-  2.0}$&$( 91.4 )$&$ 95.5  ^{+  2.2 }_{-  4.1}~^{+  1.4 }_{-  1.9 }$\\
  \multirow{2}{2cm}{  168 } & $t$ &$(143.7 )$&$159.3  ^{+  2.6 }_{-  2.9}~^{+  2.3 }_{-  2.3}$&$(149.1 )$&$152.1  ^{+  4.7 }_{-  5.9}~^{+  2.4 }_{-  2.2 }$\\
 \vspace{3pt}
      & $\bar t$ &$( 86.1 )$&$ 97.3  ^{+  1.7 }_{-  1.6}~^{+  1.3 }_{-  1.9}$&$( 87.6 )$&$ 91.7  ^{+  2.6 }_{-  3.5}~^{+  1.3 }_{-  1.8 }$\\
  \multirow{2}{2cm}{  172 } & $t$ &$(139.4 )$&$154.3  ^{+  2.9 }_{-  2.5}~^{+  2.2 }_{-  2.2}$&$(143.9 )$&$146.8  ^{+  4.3 }_{-  5.0}~^{+  2.2 }_{-  2.1 }$\\
 \vspace{3pt}
      & $\bar t$ &$( 83.4 )$&$ 94.2  ^{+  1.6 }_{-  1.5}~^{+  1.2 }_{-  1.9}$&$( 84.6 )$&$ 88.7  ^{+  2.6 }_{-  3.0}~^{+  1.3 }_{-  1.8 }$\\
  \multirow{2}{2cm}{  176 } & $t$ &$(135.2 )$&$149.6  ^{+  2.6 }_{-  2.4}~^{+  2.1 }_{-  2.1}$&$(138.9 )$&$142.7  ^{+  4.6 }_{-  6.1}~^{+  2.2 }_{-  2.0 }$\\
 \vspace{3pt}
      & $\bar t$ &$( 80.8 )$&$ 91.1  ^{+  1.6 }_{-  1.4}~^{+  1.2 }_{-  1.8}$&$( 81.4 )$&$ 85.8  ^{+  2.1 }_{-  3.5}~^{+  1.2 }_{-  1.7 }$\\
  \multirow{2}{2cm}{  180 } & $t$ &$(131.3 )$&$145.2  ^{+  2.5 }_{-  2.3}~^{+  2.0 }_{-  2.0}$&$(134.1 )$&$138.6  ^{+  5.1 }_{-  5.6}~^{+  2.1 }_{-  1.9 }$\\
 \vspace{3pt}
      & $\bar t$ &$( 78.4 )$&$ 88.2  ^{+  1.6 }_{-  1.3}~^{+  1.2 }_{-  1.8}$&$( 78.5 )$&$ 83.1  ^{+  2.1 }_{-  3.4}~^{+  1.2 }_{-  1.6 }$\\
  \multirow{2}{2cm}{  200 } & $t$ &$(113.6 )$&$125.5  ^{+  2.0 }_{-  1.9}~^{+  1.7 }_{-  1.7}$&$(112.8 )$&$119.3  ^{+  3.7 }_{-  5.4}~^{+  1.7 }_{-  1.6 }$\\
 \vspace{3pt}
      & $\bar t$ &$( 67.4 )$&$ 75.5  ^{+  1.2 }_{-  1.1}~^{+  1.0 }_{-  1.5}$&$( 65.5 )$&$ 70.9  ^{+  2.1 }_{-  3.1}~^{+  1.0 }_{-  1.4 }$\\
  \multirow{2}{2cm}{  400 } & $t$ &$( 35.1 )$&$ 38.9  ^{+  0.5 }_{-  0.2}~^{+  0.6 }_{-  0.5}$&$( 29.5 )$&$ 35.6  ^{+  2.1 }_{-  2.1}~^{+  0.5 }_{-  0.4 }$\\
 \vspace{3pt}
      & $\bar t$ &$( 19.5 )$&$ 21.7  ^{+  0.3 }_{-  0.1}~^{+  0.5 }_{-  0.6}$&$( 15.9 )$&$ 19.6  ^{+  0.9 }_{-  1.3}~^{+  0.5 }_{-  0.5 }$\\
  \multirow{2}{2cm}{  600 } & $t$ &$(14.51 )$&$16.22  ^{+ 0.19 }_{- 0.10}~^{+ 0.38 }_{- 0.30}$&$(11.01 )$&$14.42  ^{+ 0.98 }_{- 1.15}~^{+ 0.34 }_{- 0.25 }$\\
 \vspace{3pt}
      & $\bar t$ &$( 7.61 )$&$ 8.49  ^{+ 0.11 }_{- 0.05}~^{+ 0.28 }_{- 0.31}$&$( 5.58 )$&$ 7.36  ^{+ 0.48 }_{- 0.58}~^{+ 0.25 }_{- 0.26 }$\\
  \multirow{2}{2cm}{  800 } & $t$ &$( 6.97 )$&$ 7.81  ^{+ 0.08 }_{- 0.05}~^{+ 0.26 }_{- 0.21}$&$( 4.91 )$&$ 6.78  ^{+ 0.54 }_{- 0.64}~^{+ 0.22 }_{- 0.17 }$\\
 \vspace{3pt}
      & $\bar t$ &$( 3.46 )$&$ 3.88  ^{+ 0.04 }_{- 0.03}~^{+ 0.17 }_{- 0.18}$&$( 2.34 )$&$ 3.29  ^{+ 0.26 }_{- 0.31}~^{+ 0.15 }_{- 0.15 }$\\
  \multirow{2}{2cm}{ 1000 } & $t$ &$( 3.66 )$&$ 4.11  ^{+ 0.05 }_{- 0.03}~^{+ 0.18 }_{- 0.15}$&$( 2.42 )$&$ 3.50  ^{+ 0.31 }_{- 0.36}~^{+ 0.15 }_{- 0.11 }$\\
 \vspace{3pt}
      & $\bar t$ &$(1.732 )$&$1.945  ^{+0.018 }_{-0.015}~^{+0.112 }_{-0.111}$&$(1.101 )$&$1.604  ^{+0.158 }_{-0.162}~^{+0.092 }_{-0.087 }$\\
  \multirow{2}{2cm}{ 1200 } & $t$ &$( 2.05 )$&$ 2.30  ^{+ 0.03 }_{- 0.03}~^{+ 0.13 }_{- 0.10}$&$( 1.28 )$&$ 1.89  ^{+ 0.20 }_{- 0.20}~^{+ 0.10 }_{- 0.07 }$\\
 \vspace{3pt}
      & $\bar t$ &$(0.925 )$&$1.040  ^{+0.011 }_{-0.009}~^{+0.074 }_{-0.070}$&$(0.556 )$&$0.831  ^{+0.087 }_{-0.088}~^{+0.058 }_{-0.053 }$\\
  \multirow{2}{2cm}{ 1400 } & $t$ &$(1.192 )$&$1.339  ^{+0.020 }_{-0.014}~^{+0.089 }_{-0.071}$&$(0.709 )$&$1.088  ^{+0.123 }_{-0.126}~^{+0.068 }_{-0.050 }$\\
 \vspace{3pt}
      & $\bar t$ &$(0.516 )$&$0.583  ^{+0.007 }_{-0.007}~^{+0.049 }_{-0.046}$&$(0.296 )$&$0.458  ^{+0.051 }_{-0.054}~^{+0.037 }_{-0.033 }$\\
  \multirow{2}{2cm}{ 1600 } & $t$ &$(0.718 )$&$0.806  ^{+0.014 }_{-0.010}~^{+0.063 }_{-0.050}$&$(0.409 )$&$0.642  ^{+0.079 }_{-0.078}~^{+0.047 }_{-0.033 }$\\
 \vspace{3pt}
      & $\bar t$ &$(0.299 )$&$0.337  ^{+0.005 }_{-0.004}~^{+0.033 }_{-0.030}$&$(0.164 )$&$0.259  ^{+0.032 }_{-0.030}~^{+0.024 }_{-0.021 }$\\
  \multirow{2}{2cm}{ 1800 } & $t$ &$(0.443 )$&$0.497  ^{+0.010 }_{-0.008}~^{+0.045 }_{-0.035}$&$(0.242 )$&$0.389  ^{+0.050 }_{-0.049}~^{+0.032 }_{-0.023 }$\\
 \vspace{3pt}
      & $\bar t$ &$(0.177 )$&$0.200  ^{+0.004 }_{-0.003}~^{+0.023 }_{-0.020}$&$(0.093 )$&$0.151  ^{+0.019 }_{-0.018}~^{+0.016 }_{-0.014 }$\\
  \multirow{2}{2cm}{ 2000 } & $t$ &$(0.279 )$&$0.312  ^{+0.007 }_{-0.006}~^{+0.032 }_{-0.025}$&$(0.147 )$&$0.240  ^{+0.033 }_{-0.031}~^{+0.022 }_{-0.016 }$\\
 \vspace{3pt}
      & $\bar t$ &$(0.107 )$&$0.121  ^{+0.003 }_{-0.002}~^{+0.016 }_{-0.014}$&$(0.054 )$&$0.089  ^{+0.013 }_{-0.012}~^{+0.011 }_{-0.009 }$\\
 
 \bottomrule[0.08em]
  }
 { 
 \label{tab:6}
Same as Table~\protect\ref{tab:5} but for the MSTW2008 PDF set.
 }
 \end{small}

\clearpage
\section{Cross sections and uncertainties for $t  b^\prime$ and $t^\prime b^\prime$ production}\label{app:B}

\begin{small}
\renewcommand{\arraystretch}{1.1}
\TABULAR[h]{crr@{ }l}
{
\toprule[0.08em]
\multicolumn{2}{c}{{$m_{b^\prime}$ (GeV)}}&\multicolumn{2}{c}{cross section (fb)}\\
\midrule[0.05em]
  \multirow{2}{2cm}{  200             }&$b^\prime\bar t$ &$(   9394 )$&$   9556  ^{+    105 }_{-    317}~^{+    202 }_{-    203 }$\\
 \vspace{3pt}
             &$\bar b^\prime t$ &$(   5211 )$&$   5334  ^{+     41 }_{-    131}~^{+    173 }_{-    193 }$\\
  \multirow{2}{2cm}{  400             }&$b^\prime\bar t$ &$(   3256 )$&$   3607  ^{+     82 }_{-    172}~^{+    132 }_{-    120 }$\\
 \vspace{3pt}
             &$\bar b^\prime t$ &$(   1696 )$&$   1907  ^{+     18 }_{-     89}~^{+     98 }_{-    102 }$\\
  \multirow{2}{2cm}{  600             }&$b^\prime\bar t$ &$(   1457 )$&$   1738  ^{+     54 }_{-    105}~^{+     92 }_{-     79 }$\\
 \vspace{3pt}
             &$\bar b^\prime t$ &$(    713 )$&$    863  ^{+     17 }_{-     44}~^{+     64 }_{-     63 }$\\
  \multirow{2}{2cm}{  800             }&$b^\prime\bar t$ &$(    729 )$&$    920  ^{+     41 }_{-     61}~^{+     64 }_{-     54 }$\\
 \vspace{3pt}
             &$\bar b^\prime t$ &$(    338 )$&$    435  ^{+     13 }_{-     27}~^{+     41 }_{-     39 }$\\
  \multirow{2}{2cm}{ 1000             }&$b^\prime\bar t$ &$(    390 )$&$    518  ^{+     29 }_{-     37}~^{+     46 }_{-     38 }$\\
 \vspace{3pt}
             &$\bar b^\prime t$ &$(    172 )$&$    235  ^{+      9 }_{-     18}~^{+     28 }_{-     26 }$\\
  \multirow{2}{2cm}{ 1200             }&$b^\prime\bar t$ &$(    219 )$&$    307  ^{+     16 }_{-     27}~^{+     32 }_{-     26 }$\\
 \vspace{3pt}
             &$\bar b^\prime t$ &$(  92.0 )$&$ 132.1  ^{+   7.2 }_{-  10.7}~^{+  19.9 }_{-  18.3 }$\\
  \multirow{2}{2cm}{ 1400             }&$b^\prime\bar t$ &$( 126.5 )$&$ 184.6  ^{+  13.6 }_{-  16.1}~^{+  23.2 }_{-  18.4 }$\\
 \vspace{3pt}
             &$\bar b^\prime t$ &$(  51.1 )$&$  77.7  ^{+   4.4 }_{-   7.0}~^{+  13.0 }_{-  11.7 }$\\
  \multirow{2}{2cm}{ 1600             }&$b^\prime\bar t$ &$(  75.1 )$&$ 114.4  ^{+   9.5 }_{-  10.5}~^{+  16.5 }_{-  13.0 }$\\
 \vspace{3pt}
             &$\bar b^\prime t$ &$(  29.2 )$&$  46.7  ^{+   2.7 }_{-   4.6}~^{+   8.7 }_{-   7.5 }$\\
  \multirow{2}{2cm}{ 1800             }&$b^\prime\bar t$ &$(  45.5 )$&$  72.5  ^{+   6.7 }_{-   7.2}~^{+  11.9 }_{-   9.3 }$\\
 \vspace{3pt}
             &$\bar b^\prime t$ &$(  17.0 )$&$  28.5  ^{+   2.1 }_{-   2.8}~^{+   6.0 }_{-   5.1 }$\\
  \multirow{2}{2cm}{ 2000             }&$b^\prime\bar t$ &$(  28.0 )$&$  46.8  ^{+   3.9 }_{-   5.2}~^{+   8.5 }_{-   6.5 }$\\
 \vspace{3pt}
             &$\bar b^\prime t$ &$( 10.10 )$&$ 17.90  ^{+  1.39 }_{-  1.95}~^{+  4.18 }_{-  3.52 }$\\
 
\bottomrule[0.08em]
 }
{
\label{tab:h}
NLO cross sections (fb) at the LHC 14 TeV for $b^\prime \bar t$ and $\bar
b^\prime t$ as a function of $m_{b^\prime}$ obtained with the CTEQ6.6
PDF set and $V_{tb'}=1$. The first uncertainty comes from renormalisation and
factorisation scales variation and the second from PDF errors. Numbers
in parenthesis refer to the corresponding LO results.  These results
are plotted in Fig.~\ref{fig:h} where the scale and PDF uncertainties
are combined linearly.  }
\end{small}

\begin{small}
\renewcommand{\arraystretch}{1.1}
\TABULAR[h]{crr@{ }l}
{
\toprule[0.08em]
\multicolumn{2}{c}{{$m_{b^\prime}$ (GeV)}}&\multicolumn{2}{c}{cross section (fb)}\\
\midrule[0.05em]
  \multirow{2}{2cm}{  200             }&$b^\prime\bar t$ &$(   9719 )$&$   9797  ^{+    121 }_{-    315}~^{+    160 }_{-    121 }$\\
 \vspace{3pt}
             &$\bar b^\prime t$ &$(   5551 )$&$   5575  ^{+     27 }_{-    121}~^{+    133 }_{-    142 }$\\
  \multirow{2}{2cm}{  400             }&$b^\prime\bar t$ &$(   3401 )$&$   3696  ^{+     73 }_{-    184}~^{+     92 }_{-     64 }$\\
 \vspace{3pt}
             &$\bar b^\prime t$ &$(   1820 )$&$   1959  ^{+     49 }_{-     73}~^{+     66 }_{-     67 }$\\
  \multirow{2}{2cm}{  600             }&$b^\prime\bar t$ &$(   1537 )$&$   1771  ^{+     66 }_{-    104}~^{+     60 }_{-     42 }$\\
 \vspace{3pt}
             &$\bar b^\prime t$ &$(    772 )$&$    884  ^{+     31 }_{-     41}~^{+     39 }_{-     38 }$\\
  \multirow{2}{2cm}{  800             }&$b^\prime\bar t$ &$(    777 )$&$    937  ^{+     41 }_{-     65}~^{+     40 }_{-     28 }$\\
 \vspace{3pt}
             &$\bar b^\prime t$ &$(    369 )$&$    441  ^{+     15 }_{-     27}~^{+     24 }_{-     21 }$\\
  \multirow{2}{2cm}{ 1000             }&$b^\prime\bar t$ &$(    421 )$&$    526  ^{+     29 }_{-     43}~^{+     27 }_{-     20 }$\\
 \vspace{3pt}
             &$\bar b^\prime t$ &$(    189 )$&$    235  ^{+     11 }_{-     18}~^{+     16 }_{-     14 }$\\
  \multirow{2}{2cm}{ 1200             }&$b^\prime\bar t$ &$(    238 )$&$    307  ^{+     19 }_{-     28}~^{+     19 }_{-     13 }$\\
 \vspace{3pt}
             &$\bar b^\prime t$ &$( 102.0 )$&$ 131.4  ^{+   7.7 }_{-  11.3}~^{+  10.4 }_{-   9.1 }$\\
  \multirow{2}{2cm}{ 1400             }&$b^\prime\bar t$ &$( 139.1 )$&$ 183.6  ^{+  13.2 }_{-  16.8}~^{+  13.1 }_{-   9.1 }$\\
 \vspace{3pt}
             &$\bar b^\prime t$ &$(  57.0 )$&$  75.0  ^{+   5.7 }_{-   6.7}~^{+   6.9 }_{-   6.0 }$\\
  \multirow{2}{2cm}{ 1600             }&$b^\prime\bar t$ &$(  83.4 )$&$ 112.7  ^{+  10.2 }_{-  11.0}~^{+   9.1 }_{-   6.3 }$\\
 \vspace{3pt}
             &$\bar b^\prime t$ &$(  32.7 )$&$  44.1  ^{+   3.1 }_{-   4.1}~^{+   4.5 }_{-   3.8 }$\\
  \multirow{2}{2cm}{ 1800             }&$b^\prime\bar t$ &$(  51.0 )$&$  70.9  ^{+   6.1 }_{-   7.8}~^{+   6.3 }_{-   4.4 }$\\
 \vspace{3pt}
             &$\bar b^\prime t$ &$(  19.2 )$&$  26.4  ^{+   2.7 }_{-   2.5}~^{+   3.0 }_{-   2.4 }$\\
  \multirow{2}{2cm}{ 2000             }&$b^\prime\bar t$ &$(  31.7 )$&$  44.6  ^{+   4.7 }_{-   4.9}~^{+   4.4 }_{-   3.0 }$\\
 \vspace{3pt}
             &$\bar b^\prime t$ &$( 11.46 )$&$ 16.23  ^{+  1.56 }_{-  1.80}~^{+  2.08 }_{-  1.71 }$\\

\bottomrule[0.08em]
 }
{
\label{tab:i}
Same as Table~\protect\ref{tab:h} but for the MSTW2008 PDF set.
}
\end{small}

\begin{small}
\renewcommand{\arraystretch}{1.1}
\TABULAR[h]{crr@{ }lr@{ }lr@{ }l}
{
\toprule[0.08em]
\multicolumn{2}{c}{$m_{t^\prime}$ (GeV)}&\multicolumn{2}{c}{{ $m_{b^\prime}=m_{t^\prime}$}} &\multicolumn{2}{c}{{$m_{t^\prime}-m_{b^\prime}$=200 GeV}}&\multicolumn{2}{c}{{$m_{t^\prime}-m_{b^\prime}$=500 GeV}}\\
\midrule[0.05em]
  \multirow{2}{1.2cm}{  172      }&$t^\prime\bar b^\prime$ &$(  11423 )$&$  11503  ^{+    116 }_{-    359}~^{+    235 }_{-    240}$&\multicolumn{2}{c}{--}&\multicolumn{2}{c}{--}\\
 \vspace{2pt}
      &$\bar t^\prime b^\prime$ &$(   6381 )$&$   6479  ^{+     31 }_{-    150}~^{+    180 }_{-    205}$&\multicolumn{2}{c}{--}&\multicolumn{2}{c}{--}\\
  \multirow{2}{1.2cm}{  200      }&$t^\prime\bar b^\prime$ &$(   7777 )$&$   8011  ^{+    110 }_{-    313}~^{+    183 }_{-    183}$&\multicolumn{2}{c}{--}&\multicolumn{2}{c}{--}\\
 \vspace{2pt}
      &$\bar t^\prime b^\prime$ &$(   4239 )$&$   4391  ^{+     24 }_{-    124}~^{+    155 }_{-    171}$&\multicolumn{2}{c}{--}&\multicolumn{2}{c}{--}\\
  \multirow{2}{1.2cm}{  400      }&$t^\prime\bar b^\prime$ &$(    992 )$&$   1147  ^{+     27 }_{-     60}~^{+     64 }_{-     55}$&$(   2763 )$&$   3072  ^{+     74 }_{-    134}~^{+    117 }_{-    106}$&\multicolumn{2}{c}{--}\\
 \vspace{2pt}
      &$\bar t^\prime b^\prime$ &$(    470 )$&$    553  ^{+     10 }_{-     29}~^{+     42 }_{-     41}$&$(   1370 )$&$   1547  ^{+     15 }_{-     70}~^{+     86 }_{-     87}$&\multicolumn{2}{c}{--}\\
  \multirow{2}{1.2cm}{  600      }&$t^\prime\bar b^\prime$ &$(    219 )$&$    279  ^{+     10 }_{-     19}~^{+     25 }_{-     21}$&$(    465 )$&$    565  ^{+     21 }_{-     31}~^{+     41 }_{-     34}$&$(   2538 )$&$   3022  ^{+    114 }_{-    179}~^{+    138 }_{-    122 }$\\
 \vspace{2pt}
      &$\bar t^\prime b^\prime$ &$(   93.7 )$&$  121.7  ^{+    4.7 }_{-    6.3}~^{+   16.2 }_{-   15.6}$&$(  207.3 )$&$  258.2  ^{+    6.3 }_{-   14.7}~^{+   25.7 }_{-   24.3}$&$( 1207.3 )$&$ 1474.6  ^{+   34.2 }_{-   82.6}~^{+   95.4 }_{-   94.9 }$\\
  \multirow{2}{1.2cm}{  800      }&$t^\prime\bar b^\prime$ &$(   61.6 )$&$   85.5  ^{+    4.1 }_{-    7.1}~^{+   10.9 }_{-    8.6}$&$(  116.1 )$&$  154.3  ^{+    5.9 }_{-   11.2}~^{+   16.7 }_{-   13.5}$&$(  377.5 )$&$  479.3  ^{+   19.6 }_{-   30.1}~^{+   39.0 }_{-   32.2 }$\\
 \vspace{2pt}
      &$\bar t^\prime b^\prime$ &$(   24.2 )$&$   34.6  ^{+    1.3 }_{-    2.5}~^{+    5.8 }_{-    5.1}$&$(   47.3 )$&$   64.6  ^{+    2.5 }_{-    4.5}~^{+    9.5 }_{-    8.5}$&$(  163.2 )$&$  213.2  ^{+    7.1 }_{-   15.0}~^{+   23.2 }_{-   22.2 }$\\
  \multirow{2}{1.2cm}{ 1000      }&$t^\prime\bar b^\prime$ &$(   19.8 )$&$   29.7  ^{+    1.8 }_{-    2.6}~^{+    4.9 }_{-    3.8}$&$(   34.9 )$&$   50.2  ^{+    2.8 }_{-    4.3}~^{+    7.4 }_{-    5.8}$&$(   92.7 )$&$  126.7  ^{+    7.1 }_{-    9.4}~^{+   15.0 }_{-   12.0 }$\\
 \vspace{2pt}
      &$\bar t^\prime b^\prime$ &$(   7.19 )$&$  11.26  ^{+   0.66 }_{-   0.88}~^{+   2.44 }_{-   2.09}$&$(  13.15 )$&$  19.80  ^{+   0.91 }_{-   1.59}~^{+   3.80 }_{-   3.29}$&$(  36.86 )$&$  52.55  ^{+   1.43 }_{-   4.17}~^{+   8.20 }_{-   7.28 }$\\
  \multirow{2}{1.2cm}{ 1200      }&$t^\prime\bar b^\prime$ &$(   6.90 )$&$  11.21  ^{+   0.80 }_{-   1.12}~^{+   2.32 }_{-   1.76}$&$(  11.69 )$&$  18.24  ^{+   1.28 }_{-   1.69}~^{+   3.37 }_{-   2.57}$&$(  27.87 )$&$  41.23  ^{+   2.51 }_{-   3.71}~^{+   6.49 }_{-   5.05 }$\\
 \vspace{2pt}
      &$\bar t^\prime b^\prime$ &$(   2.34 )$&$   4.08  ^{+   0.24 }_{-   0.38}~^{+   1.08 }_{-   0.90}$&$(   4.09 )$&$   6.78  ^{+   0.33 }_{-   0.62}~^{+   1.63 }_{-   1.37}$&$(  10.28 )$&$  16.05  ^{+   0.60 }_{-   1.54}~^{+   3.26 }_{-   2.80 }$\\
  \multirow{2}{1.2cm}{ 1400      }&$t^\prime\bar b^\prime$ &$(   2.54 )$&$   4.47  ^{+   0.38 }_{-   0.48}~^{+   1.12 }_{-   0.84}$&$(   4.19 )$&$   7.09  ^{+   0.54 }_{-   0.72}~^{+   1.61 }_{-   1.21}$&$(   9.37 )$&$  14.91  ^{+   1.17 }_{-   1.39}~^{+   2.93 }_{-   2.24 }$\\
 \vspace{2pt}
      &$\bar t^\prime b^\prime$ &$(   0.80 )$&$   1.54  ^{+   0.10 }_{-   0.16}~^{+   0.49 }_{-   0.40}$&$(   1.37 )$&$   2.51  ^{+   0.15 }_{-   0.26}~^{+   0.73 }_{-   0.60}$&$(   3.22 )$&$   5.44  ^{+   0.39 }_{-   0.48}~^{+   1.42 }_{-   1.20 }$\\
  \multirow{2}{1.2cm}{ 1600      }&$t^\prime\bar b^\prime$ &$(   0.97 )$&$   1.86  ^{+   0.16 }_{-   0.22}~^{+   0.54 }_{-   0.40}$&$(   1.57 )$&$   2.88  ^{+   0.25 }_{-   0.31}~^{+   0.77 }_{-   0.57}$&$(   3.37 )$&$   5.83  ^{+   0.47 }_{-   0.61}~^{+   1.38 }_{-   1.04 }$\\
 \vspace{2pt}
      &$\bar t^\prime b^\prime$ &$(   0.29 )$&$   0.61  ^{+   0.05 }_{-   0.07}~^{+   0.23 }_{-   0.18}$&$(   0.48 )$&$   0.96  ^{+   0.07 }_{-   0.10}~^{+   0.34 }_{-   0.27}$&$(   1.08 )$&$   2.03  ^{+   0.14 }_{-   0.22}~^{+   0.61 }_{-   0.50 }$\\
  \multirow{2}{1.2cm}{ 1800      }&$t^\prime\bar b^\prime$ &$(   0.38 )$&$   0.79  ^{+   0.08 }_{-   0.10}~^{+   0.27 }_{-   0.19}$&$(   0.61 )$&$   1.21  ^{+   0.13 }_{-   0.14}~^{+   0.38 }_{-   0.28}$&$(   1.26 )$&$   2.37  ^{+   0.21 }_{-   0.25}~^{+   0.66 }_{-   0.49 }$\\
 \vspace{2pt}
      &$\bar t^\prime b^\prime$ &$(   0.10 )$&$   0.24  ^{+   0.03 }_{-   0.03}~^{+   0.11 }_{-   0.08}$&$(   0.17 )$&$   0.38  ^{+   0.04 }_{-   0.04}~^{+   0.15 }_{-   0.12}$&$(   0.38 )$&$   0.78  ^{+   0.07 }_{-   0.09}~^{+   0.28 }_{-   0.23 }$\\
  \multirow{2}{1.2cm}{ 2000      }&$t^\prime\bar b^\prime$ &$(   0.15 )$&$   0.34  ^{+   0.04 }_{-   0.04}~^{+   0.13 }_{-   0.09}$&$(   0.24 )$&$   0.52  ^{+   0.06 }_{-   0.07}~^{+   0.19 }_{-   0.14}$&$(   0.49 )$&$   0.99  ^{+   0.10 }_{-   0.12}~^{+   0.33 }_{-   0.24 }$\\
 \vspace{2pt}
      &$\bar t^\prime b^\prime$ &$(   0.04 )$&$   0.10  ^{+   0.01 }_{-   0.01}~^{+   0.05 }_{-   0.04}$&$(   0.06 )$&$   0.16  ^{+   0.02 }_{-   0.02}~^{+   0.08 }_{-   0.06}$&$(   0.14 )$&$   0.31  ^{+   0.03 }_{-   0.03}~^{+   0.13 }_{-   0.10 }$\\
 
\bottomrule[0.08em]
 }
{ 
\label{tab:7}
NLO cross sections (fb) at the LHC 14 TeV for $t^\prime \bar b^\prime$
and $\bar t^\prime b^\prime$ as a function of $m_{t^\prime}$ obtained
with the CTEQ6.6 PDF set and $V_{t^{\prime}b^{\prime}}=1$. The first
uncertainty comes from renormalisation and factorisation scales
variation and the second from PDF errors. Numbers in parenthesis refer
to the corresponding LO results.  These results are plotted in
Fig.~\ref{fig:7} where the scale and PDF uncertainties are combined
linearly.  }
\end{small}
\begin{small}
\renewcommand{\arraystretch}{1.1}
\TABULAR[h]{crr@{ }lr@{ }lr@{ }l}
{
\toprule[0.08em]
\multicolumn{2}{c}{$m_{t^\prime}$ (GeV)}&\multicolumn{2}{c}{{ $m_{b^\prime}=m_{t^\prime}$}} &\multicolumn{2}{c}{{$m_{t^\prime}-m_{b^\prime}$=200 GeV}}&\multicolumn{2}{c}{{$m_{t^\prime}-m_{b^\prime}$=500 GeV}}\\
\midrule[0.05em]
  \multirow{2}{1.2cm}{  172      }&$t^\prime\bar b^\prime$ &$(  11808 )$&$  11795  ^{+    115 }_{-    369}~^{+    185 }_{-    140}$&\multicolumn{2}{c}{--}&\multicolumn{2}{c}{--}\\
 \vspace{2pt}
      &$\bar t^\prime b^\prime$ &$(   6793 )$&$   6741  ^{+     48 }_{-    173}~^{+    161 }_{-    172}$&\multicolumn{2}{c}{--}&\multicolumn{2}{c}{--}\\
  \multirow{2}{1.2cm}{  200      }&$t^\prime\bar b^\prime$ &$(   8058 )$&$   8220  ^{+     76 }_{-    309}~^{+    143 }_{-    107}$&\multicolumn{2}{c}{--}&\multicolumn{2}{c}{--}\\
 \vspace{2pt}
      &$\bar t^\prime b^\prime$ &$(   4522 )$&$   4584  ^{+     29 }_{-    154}~^{+    115 }_{-    122}$&\multicolumn{2}{c}{--}&\multicolumn{2}{c}{--}\\
  \multirow{2}{1.2cm}{  400      }&$t^\prime\bar b^\prime$ &$(   1047 )$&$   1173  ^{+     24 }_{-     69}~^{+     41 }_{-     28}$&$(   2887 )$&$   3149  ^{+     58 }_{-    156}~^{+     81 }_{-     57}$&\multicolumn{2}{c}{--}\\
 \vspace{2pt}
      &$\bar t^\prime b^\prime$ &$(    510 )$&$    570  ^{+      5 }_{-     29}~^{+     28 }_{-     26}$&$(   1476 )$&$   1597  ^{+     19 }_{-     65}~^{+     56 }_{-     57}$&\multicolumn{2}{c}{--}\\
  \multirow{2}{1.2cm}{  600      }&$t^\prime\bar b^\prime$ &$(    237 )$&$    280  ^{+     11 }_{-     18}~^{+     15 }_{-     10}$&$(    497 )$&$    577  ^{+     19 }_{-     39}~^{+     26 }_{-     17}$&$(   2660 )$&$   3105  ^{+    111 }_{-    200}~^{+     92 }_{-     65 }$\\
 \vspace{2pt}
      &$\bar t^\prime b^\prime$ &$(  103.4 )$&$  121.7  ^{+    4.6 }_{-    7.7}~^{+    8.6 }_{-    7.6}$&$(  227.2 )$&$  262.1  ^{+    8.6 }_{-   16.0}~^{+   15.3 }_{-   14.1}$&$( 1306.1 )$&$ 1514.7  ^{+   49.9 }_{-   85.2}~^{+   61.7 }_{-   61.5 }$\\
  \multirow{2}{1.2cm}{  800      }&$t^\prime\bar b^\prime$ &$(   67.9 )$&$   84.3  ^{+    4.2 }_{-    7.2}~^{+    6.1 }_{-    4.1}$&$(  126.7 )$&$  154.6  ^{+    6.2 }_{-   12.6}~^{+    9.7 }_{-    6.6}$&$(  405.1 )$&$  484.2  ^{+   20.2 }_{-   34.0}~^{+   23.8 }_{-   16.3 }$\\
 \vspace{2pt}
      &$\bar t^\prime b^\prime$ &$(   27.1 )$&$   33.4  ^{+    1.7 }_{-    2.6}~^{+    3.1 }_{-    2.7}$&$(   52.6 )$&$   63.7  ^{+    2.4 }_{-    4.7}~^{+    5.2 }_{-    4.5}$&$(  179.5 )$&$  213.9  ^{+    7.1 }_{-   13.5}~^{+   13.5 }_{-   12.3 }$\\
  \multirow{2}{1.2cm}{ 1000      }&$t^\prime\bar b^\prime$ &$(   22.2 )$&$   28.7  ^{+    1.8 }_{-    2.7}~^{+    2.6 }_{-    1.8}$&$(   38.9 )$&$   49.0  ^{+    3.2 }_{-    4.2}~^{+    4.0 }_{-    2.7}$&$(  101.7 )$&$  126.2  ^{+    6.5 }_{-   10.4}~^{+    8.5 }_{-    5.8 }$\\
 \vspace{2pt}
      &$\bar t^\prime b^\prime$ &$(   8.15 )$&$  10.55  ^{+   0.60 }_{-   1.03}~^{+   1.26 }_{-   1.04}$&$(  14.82 )$&$  18.71  ^{+   0.96 }_{-   1.48}~^{+   1.96 }_{-   1.63}$&$(  41.13 )$&$  51.02  ^{+   2.51 }_{-   3.98}~^{+   4.58 }_{-   3.95 }$\\
  \multirow{2}{1.2cm}{ 1200      }&$t^\prime\bar b^\prime$ &$(   7.90 )$&$  10.56  ^{+   0.80 }_{-   1.11}~^{+   1.13 }_{-   0.76}$&$(  13.26 )$&$  17.38  ^{+   1.25 }_{-   1.71}~^{+   1.71 }_{-   1.16}$&$(  31.18 )$&$  39.97  ^{+   2.67 }_{-   3.61}~^{+   3.42 }_{-   2.33 }$\\
 \vspace{2pt}
      &$\bar t^\prime b^\prime$ &$(   2.67 )$&$   3.57  ^{+   0.23 }_{-   0.37}~^{+   0.51 }_{-   0.41}$&$(   4.66 )$&$   6.12  ^{+   0.39 }_{-   0.60}~^{+   0.79 }_{-   0.65}$&$(  11.61 )$&$  15.06  ^{+   0.60 }_{-   1.52}~^{+   1.67 }_{-   1.39 }$\\
  \multirow{2}{1.2cm}{ 1400      }&$t^\prime\bar b^\prime$ &$(   2.95 )$&$   4.04  ^{+   0.37 }_{-   0.45}~^{+   0.51 }_{-   0.35}$&$(   4.83 )$&$   6.53  ^{+   0.57 }_{-   0.70}~^{+   0.76 }_{-   0.52}$&$(  10.66 )$&$  14.12  ^{+   1.13 }_{-   1.37}~^{+   1.45 }_{-   0.99 }$\\
 \vspace{2pt}
      &$\bar t^\prime b^\prime$ &$(   0.93 )$&$   1.26  ^{+   0.12 }_{-   0.13}~^{+   0.21 }_{-   0.17}$&$(   1.57 )$&$   2.13  ^{+   0.19 }_{-   0.22}~^{+   0.33 }_{-   0.26}$&$(   3.67 )$&$   4.88  ^{+   0.28 }_{-   0.50}~^{+   0.66 }_{-   0.54 }$\\
  \multirow{2}{1.2cm}{ 1600      }&$t^\prime\bar b^\prime$ &$(   1.14 )$&$   1.61  ^{+   0.17 }_{-   0.19}~^{+   0.23 }_{-   0.16}$&$(   1.84 )$&$   2.55  ^{+   0.26 }_{-   0.29}~^{+   0.34 }_{-   0.23}$&$(   3.89 )$&$   5.32  ^{+   0.50 }_{-   0.60}~^{+   0.64 }_{-   0.44 }$\\
 \vspace{2pt}
      &$\bar t^\prime b^\prime$ &$(   0.33 )$&$   0.47  ^{+   0.04 }_{-   0.06}~^{+   0.09 }_{-   0.07}$&$(   0.55 )$&$   0.77  ^{+   0.08 }_{-   0.09}~^{+   0.14 }_{-   0.11}$&$(   1.24 )$&$   1.70  ^{+   0.15 }_{-   0.20}~^{+   0.27 }_{-   0.22 }$\\
  \multirow{2}{1.2cm}{ 1800      }&$t^\prime\bar b^\prime$ &$(   0.45 )$&$   0.65  ^{+   0.07 }_{-   0.08}~^{+   0.10 }_{-   0.07}$&$(   0.72 )$&$   1.02  ^{+   0.11 }_{-   0.12}~^{+   0.15 }_{-   0.11}$&$(   1.48 )$&$   2.08  ^{+   0.19 }_{-   0.26}~^{+   0.29 }_{-   0.20 }$\\
 \vspace{2pt}
      &$\bar t^\prime b^\prime$ &$(   0.12 )$&$   0.17  ^{+   0.02 }_{-   0.02}~^{+   0.04 }_{-   0.03}$&$(   0.20 )$&$   0.28  ^{+   0.03 }_{-   0.04}~^{+   0.06 }_{-   0.05}$&$(   0.44 )$&$   0.61  ^{+   0.07 }_{-   0.07}~^{+   0.12 }_{-   0.09 }$\\
  \multirow{2}{1.2cm}{ 2000      }&$t^\prime\bar b^\prime$ &$(   0.18 )$&$   0.27  ^{+   0.04 }_{-   0.04}~^{+   0.05 }_{-   0.03}$&$(   0.28 )$&$   0.42  ^{+   0.05 }_{-   0.05}~^{+   0.07 }_{-   0.05}$&$(   0.58 )$&$   0.83  ^{+   0.09 }_{-   0.11}~^{+   0.13 }_{-   0.09 }$\\
 \vspace{2pt}
      &$\bar t^\prime b^\prime$ &$(   0.04 )$&$   0.07  ^{+   0.01 }_{-   0.01}~^{+   0.02 }_{-   0.01}$&$(   0.07 )$&$   0.11  ^{+   0.01 }_{-   0.01}~^{+   0.03 }_{-   0.02}$&$(   0.16 )$&$   0.23  ^{+   0.02 }_{-   0.03}~^{+   0.05 }_{-   0.04 }$\\
 
\bottomrule[0.08em]
 }
{ 
\label{tab:f}
Same as Table~\protect\ref{tab:7} but for the MSTW2008 PDF set.
}
\end{small}

\begin{small}
\renewcommand{\arraystretch}{1.1}
\TABULAR[h]{crr@{ }lr@{ }l}
{
\toprule[0.08em]
\multicolumn{2}{c}{$m_{b^\prime}$ (GeV)}&\multicolumn{2}{c}{{$m_{b^\prime}-m_{t^\prime}$=200 GeV}}&\multicolumn{2}{c}{{$m_{b^\prime}-m_{t^\prime}$=500 GeV}}\\
\midrule[0.05em]
  \multirow{2}{2cm}{  400      }&$b^\prime\bar t^\prime$ &$(   1394 )$&$   1568  ^{+     10 }_{-     74}~^{+     87 }_{-     89}$&\multicolumn{2}{c}{--}\\
 \vspace{2pt}
      &$\bar b^\prime t^\prime$ &$(   2712 )$&$   3012  ^{+     65 }_{-    147}~^{+    116 }_{-    104}$&\multicolumn{2}{c}{--}\\
  \multirow{2}{2cm}{  600      }&$b^\prime\bar t^\prime$ &$(    209 )$&$    260  ^{+      6 }_{-     16}~^{+     26 }_{-     24}$&\multicolumn{2}{c}{--}\\
 \vspace{2pt}
      &$\bar b^\prime t^\prime$ &$(    462 )$&$    565  ^{+     19 }_{-     36}~^{+     41 }_{-     34}$&\multicolumn{2}{c}{--}\\
  \multirow{2}{2cm}{  800      }&$b^\prime\bar t^\prime$ &$(   47.5 )$&$   65.2  ^{+    2.3 }_{-    4.9}~^{+    9.4 }_{-    8.5}$&$(  165.5 )$&$  215.0  ^{+    5.2 }_{-   14.0}~^{+   23.1 }_{-   21.3 }$\\
 \vspace{2pt}
      &$\bar b^\prime t^\prime$ &$(  115.6 )$&$  153.7  ^{+    7.5 }_{-   10.8}~^{+   16.6 }_{-   13.4}$&$(  371.8 )$&$  472.6  ^{+   19.6 }_{-   32.7}~^{+   38.7 }_{-   31.9 }$\\
  \multirow{2}{2cm}{ 1000      }&$b^\prime\bar t^\prime$ &$(  13.18 )$&$  19.79  ^{+   1.00 }_{-   1.60}~^{+   3.78 }_{-   3.25}$&$(  37.17 )$&$  52.68  ^{+   1.66 }_{-   3.98}~^{+   8.20 }_{-   7.26 }$\\
 \vspace{2pt}
      &$\bar b^\prime t^\prime$ &$(   34.8 )$&$   50.3  ^{+    2.5 }_{-    4.3}~^{+    7.4 }_{-    5.8}$&$(   91.9 )$&$  125.6  ^{+    6.6 }_{-    9.3}~^{+   15.0 }_{-   12.0 }$\\
  \multirow{2}{2cm}{ 1200      }&$b^\prime\bar t^\prime$ &$(   4.10 )$&$   6.76  ^{+   0.39 }_{-   0.60}~^{+   1.63 }_{-   1.37}$&$(  10.33 )$&$  16.07  ^{+   0.79 }_{-   1.34}~^{+   3.18 }_{-   2.70 }$\\
 \vspace{2pt}
      &$\bar b^\prime t^\prime$ &$(  11.67 )$&$  18.15  ^{+   1.19 }_{-   1.64}~^{+   3.40 }_{-   2.60}$&$(  27.69 )$&$  40.97  ^{+   2.60 }_{-   3.49}~^{+   6.37 }_{-   4.95 }$\\
  \multirow{2}{2cm}{ 1400      }&$b^\prime\bar t^\prime$ &$(   1.37 )$&$   2.48  ^{+   0.18 }_{-   0.25}~^{+   0.94 }_{-   0.76}$&$(   3.23 )$&$   5.49  ^{+   0.39 }_{-   0.50}~^{+   1.38 }_{-   1.15 }$\\
 \vspace{2pt}
      &$\bar b^\prime t^\prime$ &$(   4.18 )$&$   7.10  ^{+   0.52 }_{-   0.74}~^{+   1.61 }_{-   1.21}$&$(   9.32 )$&$  14.85  ^{+   1.14 }_{-   1.36}~^{+   2.93 }_{-   2.24 }$\\
  \multirow{2}{2cm}{ 1600      }&$b^\prime\bar t^\prime$ &$(   0.48 )$&$   0.96  ^{+   0.08 }_{-   0.10}~^{+   0.34 }_{-   0.27}$&$(   1.08 )$&$   2.03  ^{+   0.15 }_{-   0.20}~^{+   0.62 }_{-   0.51 }$\\
 \vspace{2pt}
      &$\bar b^\prime t^\prime$ &$(   1.57 )$&$   2.88  ^{+   0.25 }_{-   0.32}~^{+   0.78 }_{-   0.57}$&$(   3.35 )$&$   5.80  ^{+   0.49 }_{-   0.59}~^{+   1.38 }_{-   1.03 }$\\
  \multirow{2}{2cm}{ 1800      }&$b^\prime\bar t^\prime$ &$(   0.17 )$&$   0.39  ^{+   0.03 }_{-   0.05}~^{+   0.16 }_{-   0.12}$&$(   0.38 )$&$   0.78  ^{+   0.07 }_{-   0.09}~^{+   0.28 }_{-   0.23 }$\\
 \vspace{2pt}
      &$\bar b^\prime t^\prime$ &$(   0.61 )$&$   1.21  ^{+   0.12 }_{-   0.14}~^{+   0.38 }_{-   0.28}$&$(   1.26 )$&$   2.37  ^{+   0.22 }_{-   0.27}~^{+   0.67 }_{-   0.49 }$\\
  \multirow{2}{2cm}{ 2000      }&$b^\prime\bar t^\prime$ &$(   0.06 )$&$   0.16  ^{+   0.02 }_{-   0.02}~^{+   0.07 }_{-   0.06}$&$(   0.14 )$&$   0.31  ^{+   0.03 }_{-   0.04}~^{+   0.15 }_{-   0.13 }$\\
 \vspace{2pt}
      &$\bar b^\prime t^\prime$ &$(   0.24 )$&$   0.52  ^{+   0.06 }_{-   0.07}~^{+   0.19 }_{-   0.14}$&$(   0.49 )$&$   0.99  ^{+   0.10 }_{-   0.12}~^{+   0.33 }_{-   0.24 }$\\
 
\bottomrule[0.08em]
 }
{
\label{tab:8}
Cross sections (fb) at the LHC 14 TeV for $b^\prime \bar t^\prime$ and
$\bar b^\prime t^\prime$ as a function of $m_{b^\prime}$ obtained with
the CTEQ6.6 PDF set and $V_{t^{\prime}b^{\prime}}=1$. The first
uncertainty comes from renormalisation and factorisation scales
variation and the second from PDF errors. Numbers in parenthesis refer
to the corresponding LO results.  These results are plotted in
Fig.~\ref{fig:8} where the scale and PDF uncertainties are combined
linearly.  }
\end{small}

\begin{small}
\renewcommand{\arraystretch}{1.1}
\TABULAR[h]{crr@{ }lr@{ }l}
{
\toprule[0.08em]
\multicolumn{2}{c}{$m_{b^\prime}$ (GeV)}&\multicolumn{2}{c}{{$m_{b^\prime}-m_{t^\prime}$=200 GeV}}&\multicolumn{2}{c}{{$m_{b^\prime}-m_{t^\prime}$=500 GeV}}\\
\midrule[0.05em]
  \multirow{2}{2cm}{  400      }&$b^\prime\bar t^\prime$ &$(   1498 )$&$   1616  ^{+     28 }_{-     58}~^{+     57 }_{-     58}$&\multicolumn{2}{c}{--}\\
 \vspace{2pt}
      &$\bar b^\prime t^\prime$ &$(   2837 )$&$   3096  ^{+     66 }_{-    160}~^{+     80 }_{-     56}$&\multicolumn{2}{c}{--}\\
  \multirow{2}{2cm}{  600      }&$b^\prime\bar t^\prime$ &$(    229 )$&$    265  ^{+      5 }_{-     16}~^{+     15 }_{-     14}$&\multicolumn{2}{c}{--}\\
 \vspace{2pt}
      &$\bar b^\prime t^\prime$ &$(    493 )$&$    572  ^{+     19 }_{-     37}~^{+     25 }_{-     17}$&\multicolumn{2}{c}{--}\\
  \multirow{2}{2cm}{  800      }&$b^\prime\bar t^\prime$ &$(   52.8 )$&$   64.0  ^{+    2.1 }_{-    4.6}~^{+    5.1 }_{-    4.4}$&$(  181.7 )$&$  217.5  ^{+    6.8 }_{-   15.8}~^{+   14.2 }_{-   13.0 }$\\
 \vspace{2pt}
      &$\bar b^\prime t^\prime$ &$(  126.2 )$&$  153.8  ^{+    6.8 }_{-   12.5}~^{+    9.7 }_{-    6.6}$&$(  399.2 )$&$  480.0  ^{+   19.9 }_{-   35.4}~^{+   23.7 }_{-   15.9 }$\\
  \multirow{2}{2cm}{ 1000      }&$b^\prime\bar t^\prime$ &$(  14.84 )$&$  18.68  ^{+   1.11 }_{-   1.56}~^{+   1.97 }_{-   1.66}$&$(  41.44 )$&$  51.09  ^{+   2.15 }_{-   3.91}~^{+   4.51 }_{-   3.89 }$\\
 \vspace{2pt}
      &$\bar b^\prime t^\prime$ &$(   38.8 )$&$   49.0  ^{+    3.2 }_{-    4.1}~^{+    4.0 }_{-    2.7}$&$(  100.8 )$&$  125.2  ^{+    6.7 }_{-   10.2}~^{+    8.4 }_{-    5.7 }$\\
  \multirow{2}{2cm}{ 1200      }&$b^\prime\bar t^\prime$ &$(   4.67 )$&$   6.07  ^{+   0.44 }_{-   0.55}~^{+   0.79 }_{-   0.65}$&$(  11.67 )$&$  15.09  ^{+   0.78 }_{-   1.50}~^{+   1.67 }_{-   1.39 }$\\
 \vspace{2pt}
      &$\bar b^\prime t^\prime$ &$(  13.24 )$&$  17.28  ^{+   1.43 }_{-   1.66}~^{+   1.71 }_{-   1.17}$&$(  31.00 )$&$  39.80  ^{+   2.57 }_{-   3.69}~^{+   3.39 }_{-   2.29 }$\\
  \multirow{2}{2cm}{ 1400      }&$b^\prime\bar t^\prime$ &$(   1.57 )$&$   2.13  ^{+   0.17 }_{-   0.22}~^{+   0.36 }_{-   0.29}$&$(   3.68 )$&$   4.89  ^{+   0.33 }_{-   0.51}~^{+   0.66 }_{-   0.54 }$\\
 \vspace{2pt}
      &$\bar b^\prime t^\prime$ &$(   4.83 )$&$   6.53  ^{+   0.60 }_{-   0.70}~^{+   0.76 }_{-   0.52}$&$(  10.62 )$&$  14.05  ^{+   1.19 }_{-   1.41}~^{+   1.45 }_{-   0.99 }$\\
  \multirow{2}{2cm}{ 1600      }&$b^\prime\bar t^\prime$ &$(   0.55 )$&$   0.76  ^{+   0.08 }_{-   0.08}~^{+   0.14 }_{-   0.11}$&$(   1.24 )$&$   1.70  ^{+   0.16 }_{-   0.18}~^{+   0.28 }_{-   0.22 }$\\
 \vspace{2pt}
      &$\bar b^\prime t^\prime$ &$(   1.83 )$&$   2.55  ^{+   0.26 }_{-   0.29}~^{+   0.34 }_{-   0.23}$&$(   3.88 )$&$   5.30  ^{+   0.48 }_{-   0.60}~^{+   0.64 }_{-   0.44 }$\\
  \multirow{2}{2cm}{ 1800      }&$b^\prime\bar t^\prime$ &$(   0.20 )$&$   0.28  ^{+   0.04 }_{-   0.03}~^{+   0.06 }_{-   0.05}$&$(   0.44 )$&$   0.62  ^{+   0.05 }_{-   0.08}~^{+   0.12 }_{-   0.09 }$\\
 \vspace{2pt}
      &$\bar b^\prime t^\prime$ &$(   0.72 )$&$   1.02  ^{+   0.11 }_{-   0.13}~^{+   0.15 }_{-   0.11}$&$(   1.48 )$&$   2.07  ^{+   0.22 }_{-   0.25}~^{+   0.29 }_{-   0.20 }$\\
  \multirow{2}{2cm}{ 2000      }&$b^\prime\bar t^\prime$ &$(   0.07 )$&$   0.11  ^{+   0.01 }_{-   0.01}~^{+   0.03 }_{-   0.02}$&$(   0.16 )$&$   0.23  ^{+   0.03 }_{-   0.03}~^{+   0.05 }_{-   0.04 }$\\
 \vspace{2pt}
      &$\bar b^\prime t^\prime$ &$(   0.28 )$&$   0.41  ^{+   0.05 }_{-   0.05}~^{+   0.07 }_{-   0.05}$&$(   0.58 )$&$   0.83  ^{+   0.10 }_{-   0.10}~^{+   0.13 }_{-   0.09 }$\\
 
\bottomrule[0.08em]
 }
{
\label{tab:g}
Same as Table~\protect\ref{tab:8} but for the MSTW2008 PDF set.
}
\end{small}

\end{document}